\documentclass[useAMS,usenatbib,letterpaper]{mn2e}
\usepackage{graphicx}
\usepackage{times}
\usepackage{color}
\def\e#1{$\cdot 10^{#1}$ }

\def\msun{M_\odot}
\def\Msun{$\msun$}
\def\Rsun{$R_\odot$}

\def\ga{\mathrel{\mathpalette\fun >}}
\def\fun#1#2{\lower3.6pt\vbox{\baselineskip0pt\lineskip.9pt
\ialign{$\mathsurround=0pt#1\hfil##\hfil$\crcr#2\crcr\sim\crcr}}}
\newcommand{\rref}[1]{\mbox{Fig.$\!$~\ref{#1}}}

\newcommand{\gsim}{\:\raisebox{.25ex}{$>$}\hspace*{-.75em}
      \raisebox{-.93ex}{$\sim$}\:}
\newcommand{\lsim}{\:\raisebox{.25ex}{$<$}\hspace*{-.75em}
      \raisebox{-.93ex}{$\sim$}\:}
\newcommand{\frat}[2]{\mbox{$\frac{\raisebox{.1ex}{$\textstyle#1$}}
{\raisebox{-.4ex}{$\textstyle#2$}}$}}
\newcommand{\fref}[1]{\mbox{Fig.$\!$~\ref{#1}}}

\newcommand{\MS}{\mbox{M$_{\textstyle \odot}$}}
\newcommand{\NMS}[1]{\mbox{$#1\,$M$_{\textstyle\odot}$}}
\newcommand{\RS}{\mbox{R$_{\textstyle\odot}$}}
\newcommand{\SN}{\mbox{SN 1987A}}
\newcommand{\dmn}[2]{\mbox{$#1\!\cdot\! 10^{#2}\,$}}
\newcommand{\ergs}{\mbox{erg$\:$s$^{-1}$}}

\title{Coupling of matter and radiation at supernova shock breakout}

\author[A.G.~Tolstov, S.I.~Blinnikov, D.K.~Nadyozhin]{A.G.~Tolstov,$^{1,2}$
S.I.~Blinnikov,$^{1,2,3}$
 D.K.~Nadyozhin,$^1$\\
    $^1$Institute for Theoretical and Experimental Physics (ITEP), \\
        Bolshaya Cheremushkinskaya, 25, 117218 Moscow, Russia, \\
$^2$Novosibirsk State University, 630090 Novosibirsk, Russia, \\
$^3$SAI, Moscow State University, 119992 Moscow, Russia
}

\date{Accepted 2012 December 6. Received 2012 November 19}

\pagerange{\pageref{firstpage}--\pageref{lastpage}} \pubyear{2012}

\def\LaTeX{L\kern-.36em\raise.3ex\hbox{a}\kern-.15em
    T\kern-.1667em\lower.7ex\hbox{E}\kern-.125emX}

\begin{document}

\label{firstpage}

\maketitle

\begin{abstract}
          Some features of the physics of radiation-dominated shock waves are discussed
          with emphasis on the peculiarities which are important for correct numerical
          modeling of shock breakouts in supernova.
          With account of those peculiarities, a number of models
          for different supernova types is constructed
          based on multigroup radiation transfer coupled to hydrodynamics.
          We describe the implementation of a new algorithm {\sc rada}, designed for modeling
          photon transfer at extremely-relativistic motions of matter, into our older code {\sc stella}.
          The results of numerical simulations of light curves, and continuum spectra are presented.
          The influence of effects of photon scattering on electrons,
          of thermalization depth and of special relativity in transfer equation is considered.
          Some cases are demonstrated, when the appearance of hard X-ray emission is
          possible at the shock breakout.
          The necessary refinements in numerical algorithms for radiative transfer and
          hydrodynamics are pointed out.
          Prospects for using the results of numerical simulation to analyze and
          interpret available and future data from space observatories are discussed.
\end{abstract}

\begin{keywords}
    supernovae: general -- ISM: jets and outflows -- radiative transfer
\end{keywords}

%________________________________________________________________
\section{Introduction}

The phenomenon of a supernova in most cases should start with a bright flash,
caused by a shock wave emerging on the surface of the star after the phase of collapse
or thermonuclear explosion in interiors.
The detection of such outbursts associated with the supernova shock breakout can be used
to obtain information about the explosion properties and presupernova parameters, which is necessary to
understand the physical processes that underlie this phenomenon.

For an accurate treatment of the shock wave propagation near the surface of a presupernova
it is necessary to perform numerical calculations which, in addition to hydrodynamics, should
account correctly for radiative transfer in moving media.
In some cases, e.g. in compact type Ib/c presupernovae, shock waves
can reach relativistic velocities \citep{Blinnikov2002,Blinnikov2003}.
Then one has to include into consideration a number of relativistic effects.
This paper presents the results of numerical simulation of shock waves
in several models for type Ib and type II supernovae, taking into account these features.

Studying the supernova shock breakout becomes
particularly topical in connection with the recent detection
of this phenomenon by the SWIFT spacecraft
\citep{Soderberg2008}.
In addition, a flash from the shock wave on the surface of red
supergiant in type II supernova SNLS-04D2dc
\citep{Schawinski2008,Gezari2008} and light echo from the shock wave of Cas A
\citep{DwekArendt2008} are detected.
Simulations \citep{Tominaga2009,Tominaga2011} with our code {\sc stella} show good agreement
with observations in case of SNLS-04D2dc.
The enormous luminosity of SN~2006gy is most successfully explained by models
where the radiative shock wave provides almost all radiation for many months
\citep*{Woosley2007}.
There is a possibility of obtaining new data if the experiments similar to LOBSTER
space observatory \citep{CalzavaraMatzner2004}, the satellite EXIST
\citep[Energetic X-ray Imaging Survey Telescope, see][]{Grindlay2003,Band2008}
or any X-ray station of a similar type would be launched in the future.
E.g., the experiment MAXI (Monitor of All-sky X-ray Image) on board the module Kibo at ISS
\citep{Matsuoka1997} is already started.

In this paper we compare the results of the predictions for shock breakout
for three codes: the equilibrium diffusion gray radiation-hydro code {\sc snv},
nonequilibrium multigroup radiation-hydro  code {\sc stella} and ultrarelativistic nonequilibrium multigroup
transfer code {\sc rada}.
We check the sensitivity of the predictions of the shock emission
to the parameters of the numerical scheme,
such as the boundaries of the frequency interval (including the X-ray range),
and to the approximations in the opacity description.
It is found that high-temperature peak behind the shock front and production
of hard ``tail'' in spectrum are suppressed by
extremely low true absorption, with a cross-section at the level $10^{-6}$ of Thomson
scattering in a presupernova SN~Ib.
This level of absorption can be provided by double Compton effect
\citep{MandlSkyrme1952,Weaver1976,Lightman1981},
so it must be taken into account in realistic models of radiation-dominated shock
waves in a supernova.
Some additional refinements necessary
in the methods of constructing shock-breakout models are pointed out. Among them
%Also in this paper%, using the model for type Ib supernova and a model for SN 1987A as an example, \
we consider how the relativistic and geometric effects in radiative transfer in a
comoving frame of reference influence the predictions of
supernova light curves and spectra at the epoch of shock-breakout.

%________________________________________________________________
\section{Radiation dominated shock waves}

Let us consider some of the peculiarities of a shock wave
when one cannot neglect radiation, and, moreover,
when the shock front structure is determined by energy and momentum of photons.
This situation is typical for supernova explosions,
therefore, its description is important for building adequate models.

There are many theoretical articles and several books where properties
of radiation-dominated shock are considered. In most theoretical works
it is supposed that the medium is optically thick, i.e. photons born in the shock
waves are absorbed by the cold matter upstream.
This pictures holds in a stellar explosion before the actual outburst of supernova light
while the shock is still buried under the photosphere.

{\color{black}
A brief review of the contributions that are important for the history of the problem
is given by \citet{BlinnikovTolstov2011}. Below we mention only the papers most important for further discussion.
}

A very detailed study of shock structure with radiation has been performed by \citet{Weaver1976}:
he has shown that production of observable hard (gamma-ray) radiation
at supernova shock breakout is very problematic.
Nevertheless, \citet{BlandfordPayne1981a,BlandfordPayne1981b} have predicted the possibility of
a nonthermal ``tail'' in spectra born in shock waves in hot media.
However, these authors have neglected a number of effects associated with Compton
scattering taken into account by \citet{Weaver1976}.
Further refinement of the theory with regard to the thermal Compton effect has been made by
\citet{LyubarskiiSyunyaev1982}, \citet*{Fukue1985}, \citet{Becker1988}
%, \citet{Riffert1988}, \citet{NagirnerPoutanen1993}, \citet{PsaltisLamb1997,PsaltisLamb1999}
 and other researchers.

Discontinuities inevitably develop in gas parameters in a shock wave which is not too weak
and not too strong (while radiation energy and momentum can be neglected), if one takes into
account only thermal conduction and neglects viscosity \citep[see section VII.3 in][]{ZeldovichRaizer1966}.
In real gases the discontinuities are smoothed by viscosity which transforms the ordered motion
(kinetic energy of the flow) into the heat (chaotic motion of particles) on the distance of the order
of one mean free path of particles.
This is called a viscous jump \citep{ZeldovichRaizer1966,Shu1992}.
At very high amplitudes of the shock, when the energy density and pressure of radiation
become large in comparison with those of matter, the situation is
different and the viscous jump disappears.
Gas goes from the initial state to the final one under the action of
radiative heat conduction, even if the viscosity of matter is not taken into account
{\color{black}
\citep{Belokon1959}}. Such a situation is achieved
when radiation pressure $P_r$ and the gas pressure $P_g$ behind the shock front are related as
$P_r/P_g\simeq 4.4$.

\citet{ImshennikMorozov1964} analysed various options for transfer equation
(simple diffusion equation and the Eddington approximation in the moving medium).
Plasma upstream the shock front was taken to be cold ($T_0=0, \; P_g=0$).
Their solution is more rigorous than presented by \citep{Belokon1959} and it yields
$P_r/P_g\simeq 8.5$ for the condition of disappearance of viscous jump.
The compression in the shock should then exceed a critical value $\rho_1/\rho_0 = 6.68$.
However, see \citet{WeaverChapline1974}, where it is again obtained $P_r/P_g \simeq 4.4$.
It should be noted that in determining the structure of the shock front,
the flux of photon momentum, i.e. the radiation pressure, begins to play a role
comparable to the flux of energy.
Such conditions are often achieved in astrophysical shock waves, in particular, in supernovae.

The viscous density jump disappears during propagation of the shock wave inside of a presupernova
if radiation pressure  is appreciably higher than the plasma pressure and when
there is a fairly intense processes of absorption and creation of photons, not just
their pure monochromatic scattering.
Note that \citet{ImshennikMorozov1964} assumed extinction equal to the true absorption,
scattering was not taken into account.

Calculations of \citet{Weaver1976} have been performed
in the diffusion approximation with a simplified description of Compton scattering
and various processes of absorption and creation of photons, and they have resulted
in a critical ratio $P_r/P_g \simeq 4.45$.
He assumed that the radiation has an equilibrium blackbody spectrum.
This was criticized by \citet{BlandfordPayne1981a,BlandfordPayne1981b, Riffert1988, Becker1988}.
In our work, we make no assumptions about the equilibrium of radiation, and take into account
in the calculations of the evolution of photons the same terms as
\citet{BlandfordPayne1981a,BlandfordPayne1981b}.
Nevertheless, our results are closer to the results of \citet{Weaver1976}, than to the results of his
critics: we show that the achievement of high temperatures in the radiation-dominated
shock waves is unlikely with a realistic allowance for the photon production processes.

An important parameter is the optical depth $\tau_k$ of the heating zone for strong shocks.
In the approximation of \citet{ImshennikMorozov1964},
{\color{black}
\begin{equation}
 \tau_k \sim {12c \over 7D_s} \; ,
\label{taukIM}
\end{equation}
where $D_s$ is the shock front velocity and $c$ is the speed of light.
}
The approximate solution by \citet{Klimishin1968} uses the concept of a critical temperature $T_k$,
introduced by \citet{Raizer1957}.
The solution by \citet{Klimishin1968} is closer to the conditions of supernovae
than the solution by \citet{ImshennikMorozov1964},
because $T_k$ in supernovae is low at a low density. It gives
{\color{black}
\begin{equation}
 \tau_k \sim {14c \over 3D_s} ,
\label{taukKlim}
\end{equation}	 
}
i.e., a value of the same order of magnitude.

Thus, the shock wave propagates without energy losses as
long as the lower limit on the optical depth holds
\citep{Ohyama1963,ImshennikMorozov1964,Chevalier1981,ImshennikNadezhin1989}:
{\color{black}
\begin{equation}
\tau = \frac{\delta R}{l} \ga \frac{c}{D_s} \; ,
\label{shf}
\end{equation}
where $\delta R$ is the distance from the shock front to the photosphere,
$l$ is the photon mean free path.
}
Estimate (\ref{shf}) was actually obtained by \citet{Sachs1946} as the thickness of the
radiation-dominated shock front instead of the more accurate estimates (\ref{taukIM}) and (\ref{taukKlim}).
To obtain an estimate (\ref{shf}), the shock front travel time $t_{\delta R} = \delta R/D_s$
 to the distance $\delta R$ should be set equal to $t_d = (\delta R)^2/cl$ ---
the photon diffusion time to the same distance.

As the shock approaches the photosphere, relation (\ref{shf}) breaks down
and an outburst occurs at the stellar surface.
The shock propagation in this regime can no longer be considered adiabatic,
which makes it difficult to construct analytical solutions
and necessitates numerical calculations of a process in which
radiative transport plays a very important role.

%__________________________________________________________________

\section{Self-similar and numerical solutions of supernova shock breakout}
\subsection{Self-similar solutions}

The problem of shock breakout in decreasing density
was first formulated by \citet{GandelmanFrankKamenetskii1956}.
They found a self-similar solution for a shock wave approaching
stellar surface down the density decreasing as a power of distance to
the surface. The properties of the solution for various values
of the power and adiabatic index were calculated by \citet{Grasberg1981}.
The solution was confirmed by \citet{Sakurai1960} who
showed in addition that subsequent expansion in a rarefaction wave
can be described by another self-similar solution.
The bulk of the pressure created by the shock before emergence accelerates
matter in the rarefaction wave by a factor of $\approx (1.5 - 2)$
(see details in \citet{LitvinovaNadezhin1990}).

As the shock approaches the surface where
the presupernova matter density $\rho$ falls sharply,
it accelerates following a self-similar solution.
A good overview of analytical solutions and estimates for this problem was given by
\citet{ImshennikNadezhin1989}; newer analytical results were
presented by \citet{MatznerMcKee1999}.
\citet*{JohnsonMcKee1971,Colgate1972,McKeeColgate1973} studied the shock acceleration
to relativistic velocities in compact presupernovae.
Previously, it was assumed that this could generate
an outburst of X-ray or even gamma-ray radiation \citep{Colgate1974,BisnovatyiKogan1975}.
The difficulty of the gamma-ray photon production in this scenario was shown by \citet{Weaver1976}.

\subsection{Numerical algorithm {\sc stella}}

Very few detailed calculations of shock breakout
have been published. Previously, they were made
by invoking a number of rough approximations,
such as the use of constant Eddington factors,
the single-group approximation, and the neglect of
the expansion effect in opacity
\citep{KleinChevalier1978,EnsmanBurrows1992,KellyKorevaar1995}.
{{\color{black}
Here, we use the code {\sc stella}
\citep{Blinnikov1998,Blinnikov2006}, which is designed
to solve the problem of the radiative transfer of
nonequilibrium radiation with allowance made for
hydrodynamics and, subsequently, to model the light
curves of supernovae. 
{\sc stella} is the hydrodynamics code
that incorporates multigroup radiative transfer. 
We have a very fine grid in outer layers and there are always several mesh points within the shock front. The absence of viscous jump in radiation-dominated shocks is vital for computing their hydrodynamics with accuracy of the computation of radiative transfer. Thus, the use of artificial viscosity is not necessary for calculation of the radiation-dominated shocks in outer layers.

The time-dependent equations are solved implicitly for the angular
moments of intensity averaged over fixed frequency bands.
The number of frequency groups available for current
workstation computing power, typically 100-300 in the range from $1$\AA\,to $5\cdot 10^4$\AA, is adequate to
represent, with reasonable accuracy, the nonequilibrium
continuum radiation. {\sc stella}  includes in full opacity photoionization, free-free
absorption, lines and electron scattering. The equation of state treats the ionization by
equilibrium Saha's approximation.
}
Thus, we use the method
of complete multigroup radiation hydrodynamics in
which the defects of older approaches were corrected.
This method is well applicable for the models of
type II supernovae, such as SN 1987A and SN 1993J \citep{Blinnikov1998,Blinnikov2000}.
{\color{black}
Our {\sc stella} method provides the most
reliable predictions for an outburst to be made as long as
the matter velocity $u$ is less than
$\sim (20-30)$~\% of the speed of light $c$. The method consistently takes into account all the terms of
the order of $u/c$ in hydrodynamic and radiation transfer equations whereas the terms of the order of $(u/c)^2$ are neglected.

 \subsection{Numerical algorithm {\sc rada}}
If the velocity of matter behind the shock wave reaches a significant fraction of the speed of light,
$\beta\equiv u/c\ga (0.2-0.3)$, then it is more relevant for {\sc stella} to use
an algorithm {\sc rada},
which is able to solve the radiative transfer equation in comoving frame
up to the values of the Lorentz factor,
$\gamma \equiv (1-\beta^2)^{-1/2}\sim 1000$. The use of {\sc rada} is more relevant for 
relativistic flows in the most energetic supernovae and GRB afterglows, but in this paper we would like to use {\sc stella} for testing
{\sc rada} algorithm and find out the corrections to {\sc stella} algorithm at the limit of 
applicability.
}  
The transport equation solved by {\sc rada} looks like this \citep{Mihalas1980}:
\begin{eqnarray}
\label{main}
\frac{\gamma}{c}(1+\beta\mu)\frac{\partial I(\mu,\nu)}{\partial t}+
\gamma(\mu+\beta)\frac{\partial I(\mu,\nu)}{\partial r}
+\gamma(1-\mu^2) \times \nonumber \\
\times \Big[\frac{(1+\beta\mu)}{r}-
\frac{\gamma^2}{c}(1+\beta\mu)\frac{\partial \beta}{\partial t}-
\gamma^2(\mu+\beta)
\frac{\partial\beta}{\partial r}\Big] \times  \nonumber \\
\times \frac{\partial I(\mu,\nu)}{\partial\mu}
-\gamma\Big[\frac{\beta(1-\mu^2)}{r}+
\frac{\gamma^2}{c}(1+\beta\mu)\frac{\partial \beta}{\partial t}+ \nonumber \\
+\gamma^2\mu(\mu+\beta)
\frac{\partial\beta}{\partial r}\Big]
\nu\frac{\partial I(\mu,\nu)}{\partial\nu}
+3\gamma\Big[\frac{\beta(1-\mu^2)}{r}+  \nonumber \\
+\frac{\gamma^2\mu}{c}(1+\beta\mu)\frac{\partial \beta}{\partial t}
+\gamma^2\mu(\mu+\beta)
\frac{\partial\beta}{\partial r}\Big]I(\mu,\nu) =  \nonumber \\
=\eta(\nu)-\chi(\nu)I(\mu,\nu) \; .%\nonumber
 \label{mihtraneq}
\end{eqnarray}
Here, $\eta$ -- emission coefficient, $\chi$ -- absorption coefficient, $\mu$ --
cosine of the angle between the photon momentum and the radial direction,
All values refer to the comoving frame of reference.
We present here the equation in order to give the reader the scale of its complexity
and non-triviality of its numerical solution.

To simultaneously solve the relativistic radiative
transfer equation in a comoving frame of reference
with the hydrodynamic equations, we must determine
the coupling, that is the quantities appearing both in the transfer
and in hydrodynamics. For example, the radiation
moments $J_{\nu}$,$H_{\nu}$,$K_{\nu}$ can act as these quantities.
They should be found in the calculations at hydrodynamical
grid points ${r_i,\nu_m}$, i.e., for each energy group
${\nu_m}$ and each Lagrangean zone ${r_i}$. Thus, the problem
is reduced to finding the radiation moments on the
radius-frequency grid ${r_i,\nu_m}$ at each instant of time $t_k$
in the outer stellar layers where the optical depth is
small. To calculate the moments, we must know the
radiation intensity $I_{i,k,m}$ as a function of $\mu$.

 Here, we use a characteristic method of
solving the radiative transfer equation developed previously
\citep{TolstovBlinnikov2003}. This allows the
radiation intensity to be found at a given grid point
for some set of cosines $\mu_l$. The set of  $\mu_l$ is chosen by
optimizing the calculation of the first moment $J_{\nu}$ with
a specified accuracy by the number of points $\mu_l$ at a
given grid point ${r_i,\nu_m}$.

The initial intensities at the first step $t_K$ are chosen
under the assumption of blackbody radiation and at
each successive step starting from $t_{K+1}$ are the result
of a linear interpolation of the intensities calculated at
the previous step. In outer stellar layers the optical
depth is small, but the radiation becomes increasingly
close to blackbody spectrum as one goes deeper
into the star. Therefore, the radiative transfer equation
is numerically solved only in the outer layers (as a
rule, several tens of radial mesh points) and the radiation in
deeper layers is assumed to be blackbody.

For each grid point ${r_i,\nu_m}$, we solve the relativistic
radiative transfer equation by the method of short
characteristics. The characteristic emanates from
point ${r_i,\mu,\nu_m,t_k}$ and ends at time $t_{k+1}$, making
some number of steps ${s_n}, n=1...N$. In our
calculations, we assume a linear change in emission
and absorption coefficients $\eta$ and $\chi$ in the interval
of the characteristic $ds$ due to different times at the
boundaries of this interval:
\begin{equation}
    \frac{dI(s)}{ds} = \eta + \widetilde{\eta} s + \chi I(s)+
    \widetilde{\chi} s I(s)\, ,
\end{equation}
where $\eta, \chi, \widetilde{\eta}, \widetilde{\chi}$ are constants.
The analytical solution of this equation is:
\begin{equation}
    I(s) = \Big(I_0 - \frac{\widetilde{\eta}}{\widetilde{\chi}}\Big)
    e^{-\tau}+ \frac{\widetilde{\eta}}{\widetilde{\chi}} +
    \Big(\eta-\chi\frac{\widetilde{\eta}}{\widetilde{\chi}}\Big)
    e^{-\tau}\int_0^s e^{\tau(s')} ds'
\label{Inten}
\end{equation}
where $\tau = 0.5 \widetilde{\chi}s^2+\chi s$.

Apart from the radiative transfer equation in a
comoving frame of reference, {\sc rada} takes into account
the differences in delay of the radiation from the supernova
explosion. This stems from the fact that the radiation
from the edge of the star visible to an observer comes
later than that from the central regions.

{\color{black}
Solving radiative transfer equation (\ref{mihtraneq}) is important for allowance for
time delay effect in the most energetic supernovae due to relativistic corrections.
Allowance for this can not be performed by {\sc stella} algorithm which solves radiative transfer by moment equations.
}

The radiation flux $F_{\nu}$ at some time $t_{\mathrm{obs}}$ at the point
of observation is an integral over all visible points of a
stellar surface radiating with some intensity $I$
determined by the solution of the transfer equation:

%\begin{eqnarray}
%    \label{far_transport}
%  &&  F_{\nu}(t_{\mathrm{obs}})=2\pi\int_{\mu_{\mathrm{min}}}^{1}
%    \mu p^2 \times \nonumber \\
%    \quad
%  &&  \times I_0(R(\mu),\nu\Big(\frac{\nu_0}{\nu}\Big),\cos\delta_0(\cos\delta))
%    \Big(\frac{\nu}{\nu_0}\Big)^3 d\mu
%\end{eqnarray}

\begin{equation}
    \label{far_transport}
F_{\nu}(t_{\mathrm{obs}})=\frac{2\pi}{D^2}\int_{\mu_{\mathrm{min}}}^{1}
    \mu R^2(t)\, I[t,\nu,\mu,R(t)]\, d\mu\, .
\end{equation}
%Here, $\mu$ is the cosine of the angle between the normal
%to the radiation surface and the observer's direction,
%$\nu$ is the radiation frequency, $R(\mu)/D=p(\mu)$,
%$\cos\delta=(\mu-p(\mu))/l(\mu)$,
%$l(\mu)=(1+p^2(\mu)-2p(\mu)\mu)^{1/2}$, $D$
%is the distance between the object under consideration
%and the remote observer, and the subscript $0$
%refers to the quantities in the comoving frame of reference.
Here, $\nu$ is the radiation frequency, $D$
is the distance of the star from a remote observer,
$\mu$ is the cosine of the angle between the normal
to the radiating surface and the line along observer's direction
that intersects the stellar surface at local time $t$ and at radius $R(t)$.

The observer's time $t_{\mathrm{obs}}$ is connected with $t$
by the relation
\begin{equation}\label{tobst}
    t_{\mathrm{obs}}= t +\frac{D}{c} -\frac{R}{c}\,\mu\, .
\end{equation}
Equations (\ref{far_transport},\ref{tobst}) are written in a reference frame
at rest. The quantities $I$, $\nu$, $\mu$ therein are connected by standard
transformations with their values provided by {\sc rada} in comoving frame.

The light curve at the
epoch of supernova shock breakout could be affected significantly
in case of large and highly accelerated envelopes (see details in Appendix B).
We will demonstrate examples of this in the next sections.

\section{Shock breakout in supernova models}

We have already presented some of the results on the
first outburst of hard radiation generated by the shock
breaking out from the presupernova for the models
of SN~1993J \citep{Blinnikov1998}
and for SN~Ib \citep{Blinnikov2002,Blinnikov2003}.
Some of the results were only
reported at conferences by S.I.~Blinnikov in 1997%\citep{Blinnikov1997}
\footnote{%Supernova Explosions: Their Causes and Consequences, Institute for Theoretical %Physics, Santa Barbara, August 5-9, 1997  (
http://online.kitp.ucsb.edu/online/supernova/snovaetrans.html
%)
}.
Below, we publish these results and provide additional information
about the outbursts at shock breakout in SN~II-P
(in particular, the SN~1987A type) and SN~Ib
and consider the dependence of outburst properties and shock breakout
hydrodynamics on presupernova parameters.

%__________________________________________________________________
%DKN

\subsection{Shock wave breakout calculated by {\sc snv} code}

 Some properties of the shock wave breakout as calculated by
 D.K.~Nadyozhin in 1993 are gathered in Table$\:$\ref{intr}.
 The calculations were done in the approximation of the radiation
 heat conductivity (that is equilibrium diffusion of radiation) with the aid of the hydrodynamic code {\sc snv} that
 was used in previous work of 
 \citet{GrassbergImshennik1971} and 
 \citet{ImshennikNadezhin1989}.
 All the presupernova models are due to \citet{WeaverWoosley1993}.

\begin{table*}
\begin{minipage}{94mm}
 \caption{The intrinsic properties of the light curves
           during SW breakout.}
 \label{intr}
\vspace*{2mm}
\begin{tabular}{||lllccll||}\hline\hline
  & & & & & & \\[-2mm]
 Model & \multicolumn{1}{c}{M} & \multicolumn{1}{c}{$R$}
        & $E_{\rm exp}$ & $T_p$ & $\Delta t_{p3}$ & $L_p$ \\
       & \multicolumn{1}{c}{(\MS)}& (\RS)& ($10^{51}$erg) & ($10^5$K) &
 (sec) & \\[2mm] \hline
 & & & & & &  \\[-2mm]
 SN II  & 11.06 & 335 & 1.3 & 2.94 & 1000 & 2.92\\
 SN II  & 15.08 & 496 & 1.3 & 2.45 & 1100 & 3.08 \\
 SN II  & 15.08 & 496 & 1.8 & 2.70 & 720 & 5.01 \\
 SN II  & 25.14 & 881 & 1.3 & 1.53 & 7200 & 1.48 \\
 SN II  & 35.19 & 1160 & 1.3 & 1.31 & 6900 & 1.37 \\
 SN 1993J & 3.81 & 629 & 1.3 & 2.70 & 1900 & 7.24 \\[-1.3mm]
 (j13a7) & & & & & &  \\
 SN Ib & 3.51211 & 0.763 & 1.3 & 51.0 & 0.12 & 1.45 \\[-1.3mm]
 truncated & & & & & & \\[-1.3mm]
 at zone 327 & & & & & & \\
 SN Ib & 3.51250 & 1.23 & 1.3 & 35.0 & 0.60 & 0.807 \\[-1.3mm]
 truncated & & & & & & \\[-1.3mm]
 at zone 346 & & & & & & \\
 \SN\ & 22.12 & 66.6 & 1.3 & 6.01 & 90 & 2.07 \\
 \SN\ & 18.10 & 37.8 & 1.3 & 7.47 & 33 & 1.61 \\[2mm]
  \hline\hline\\[-2mm]
\end{tabular}
\medskip
 M and $R$ --- presupernova mass and radius, respectively\\
          $E_{\rm exp}$ --- the kinetic energy at infinity\\
          $T_p$ --- the peak temperature\\
          $\Delta t_{p3}$ --- the width of the light curve peak at
     $T=\frac{1}{2}\, T_p$ (3 stellar magnitudes below the peak luminosity)\\
          $L_p$ --- the peak luminosity (in $10^{45}$\ergs)\\
\end{minipage}
\end{table*}

 The data for six models from Table$\:$\ref{intr} were additionally
 worked up to take into account the time-delay spread of the light
 curves.
 The algorithm of the corresponding filtering procedure is described
 in Appendix A.
 The results are compiled in Table$\:$\ref{insp} and shown
 in Figures \ref{M11}--\ref{87AM18} below.
 In all the figures the dashed
 lines correspond to the time-spread light curves while the solid
 ones show the case when the time-delay effect is neglected.
 In order to make the light curves discernible both near the peak
 and in their tail part, the horizontal time axes are shown in terms
 of $\log (t-t_0)$ where $t_0$ is the time the SW takes
 to reach stellar surface.

\begin{table*}
\begin{minipage}{90mm}
 \caption{The comparison of the intrinsic and time-spread light curves.}
 \label{insp}
\vspace*{2mm}
\begin{tabular}{||lllllll||}\hline\hline
   & & & & & & \\[-2mm]
 \multicolumn{1}{||c}{Model} & \multicolumn{1}{c}{M} &
 \multicolumn{1}{c}{$R/c$} & \multicolumn{1}{c}{$L_p$}
  & \multicolumn{1}{c}{$L_{ps}$}
  & $\Delta t_{p1}$ & $\Delta t_{p1s}$ \\
  & \multicolumn{1}{c}{(\MS)}& \multicolumn{1}{c}{(sec)}
  & \multicolumn{2}{c}{$\left(10^{45}\ergs\right)$}
  & (sec) & (sec)\\[2mm] \hline
  & & & & & & \\[-2mm]
 SN II  & 11.06 & 779 & 2.92 & 1.09 & 200 & 672\\
 SN II  & 15.08 & 1150 & 3.08 & 1.27 & 343 & 991\\
 SN II  & 35.19 & 2690 & 1.37 & 1.07 & 2390 & 3230\\
 SN 1993J & 3.81 & 1460 & 7.24 & 3.20 & 462 & 1380\\
 SN Ib & 3.51211 & 1.77 & 1.45 & 0.056 & 0.028 & 1.42\\
 \SN\ & 18.10 & 87.8 & 1.61 & 0.358 & 12.2 & 59.4 \\[2mm]
       \hline\hline\\[-2mm]
       \end{tabular}
\medskip
Here $L_p$ and $L_{ps}$ are the peak intrinsic and
            time-spread luminosities,
            $\Delta t_{p1}$ and $\Delta t_{p1s}$ are the widths of the
            light curves at 1 stellar magnitude below $L_p$ and $L_{ps}$,
            respectively.
            For all the supernova models in Table$\:$\ref{insp}, $E_{\rm exp}=\dmn{1.3}{51}$erg.
\end{minipage}
\end{table*}

\begin{figure}
\begin{center}
\includegraphics[width=84mm]{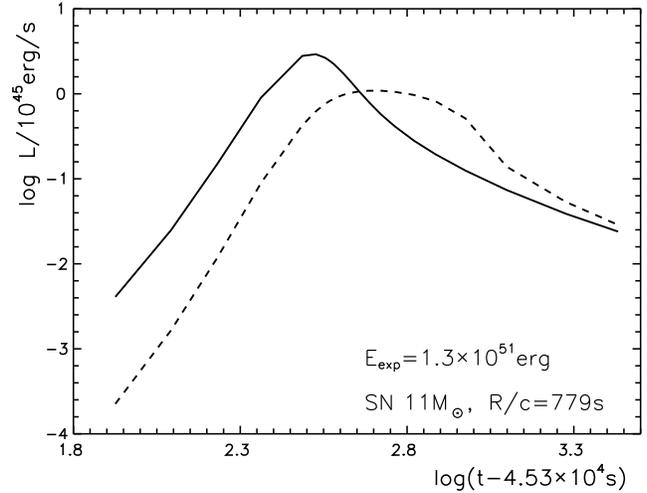}
\caption{The model SN$\,$II \NMS{11.06}.
\label{M11}
}
\end{center}
\end{figure}

\begin{figure}
\begin{center}
\includegraphics[width=84mm]{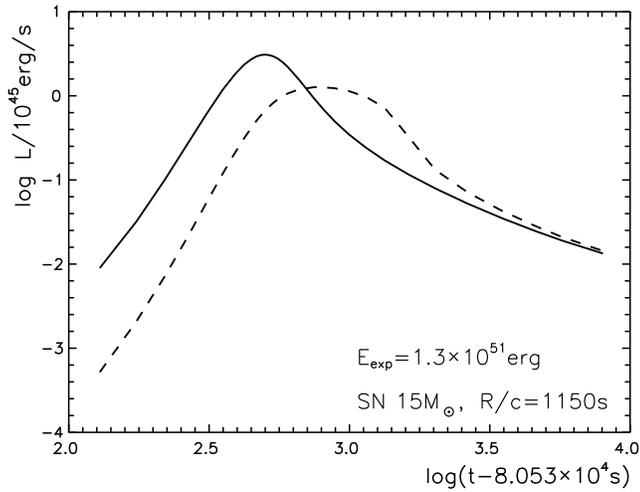}
\caption{The model SN$\,$II \NMS{15.08}.
\label{M15}
}
\end{center}
\end{figure}

\begin{figure}
\begin{center}
\includegraphics[width=84mm]{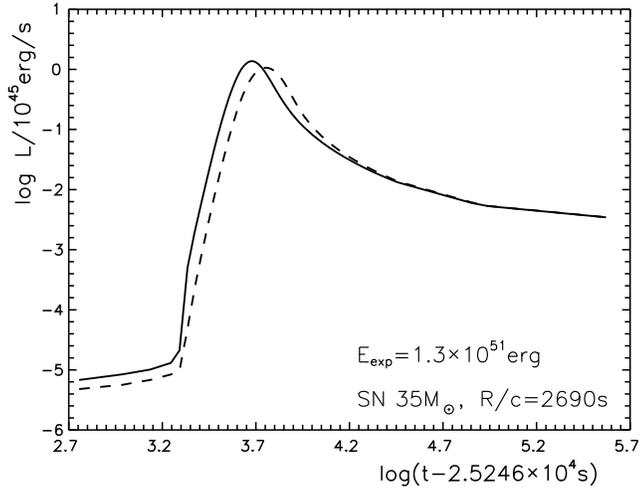}
\caption{The model SN$\,$II \NMS{35.19}.
\label{M35}
}
\end{center}
\end{figure}

\begin{figure}
\begin{center}
\includegraphics[width=84mm]{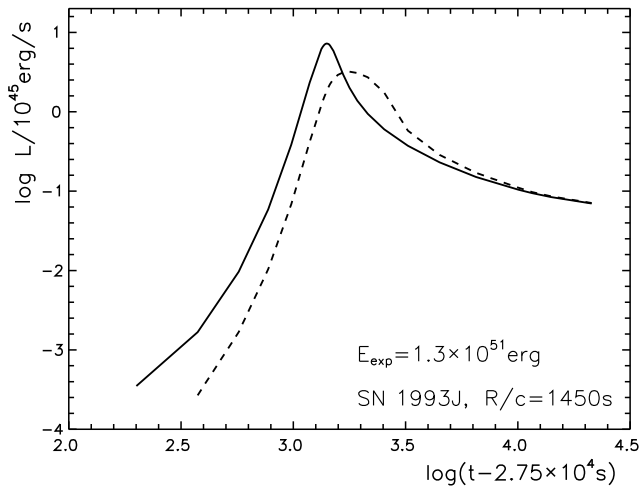}
\caption{The model SN$\,$1993J \NMS{3.81}.
\label{SN93J}
}
\end{center}
\end{figure}

\begin{figure}
\begin{center}
\includegraphics[width=84mm]{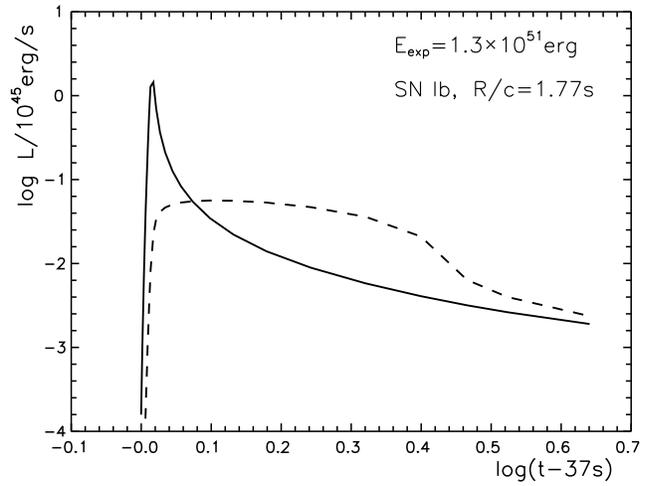}
\caption{The model SN$\,$Ib \NMS{3.51211}.
\label{SNIb07}
}
\end{center}
\end{figure}

\begin{figure}
\begin{center}
\includegraphics[width=84mm]{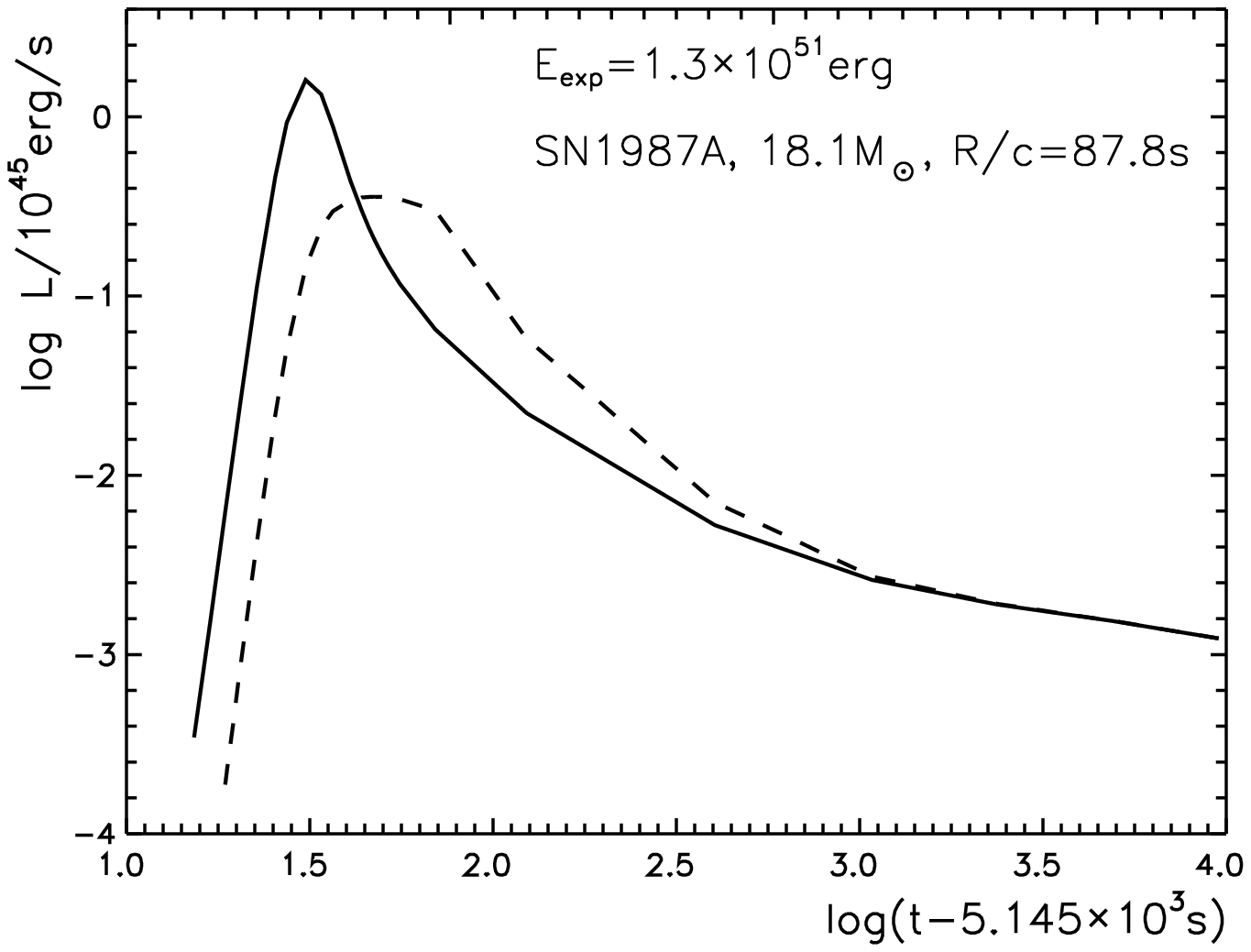}
\caption{The model \protect\SN\ \NMS{18.10}.
\label{87AM18}
}
\end{center}
\end{figure}

 Figure$\,$\ref{spIb} shows the spectral luminosities for
 the SN$\,$Ib model at 5 different points of time (as usual
 the dashed curves correspond to the time-spread case).
 The solid curves are nothing else but the Planck distributions
 at given effective temperature $T(t)$. The time $t=1.04$s corresponds
 to the luminosity peak with $T(t)=T_p=\dmn{5.1}{6}$K
 (Table$\,$\ref{intr}) and the spectrum attains its maximum
 at $h\nu = 2.82\, kT =1.24\,$keV. The dashed curves represent
 the result of the superposition of the Planckian spectra
 with the temperature $T(t)$ varying in
 the time ``window" $[t-R/c,\: t]$. Consequently, the time-spread
 spectral luminosity keeps a good admixture of the keV photons
 during $\Delta t\approx R/c$ that is much longer than the intrinsic
 luminosity does (compare the dashed and solid curves at $t=2.51\,$s).
 With time, the difference between the dashed and solid curves disappears:
 at $t=11.2\,$s they become virtually indistinguishable.
 Note that the total (time-integrated) spectral luminosity is
 exactly the same for both the cases.

\begin{figure}
\begin{center}
\includegraphics[width=84mm]{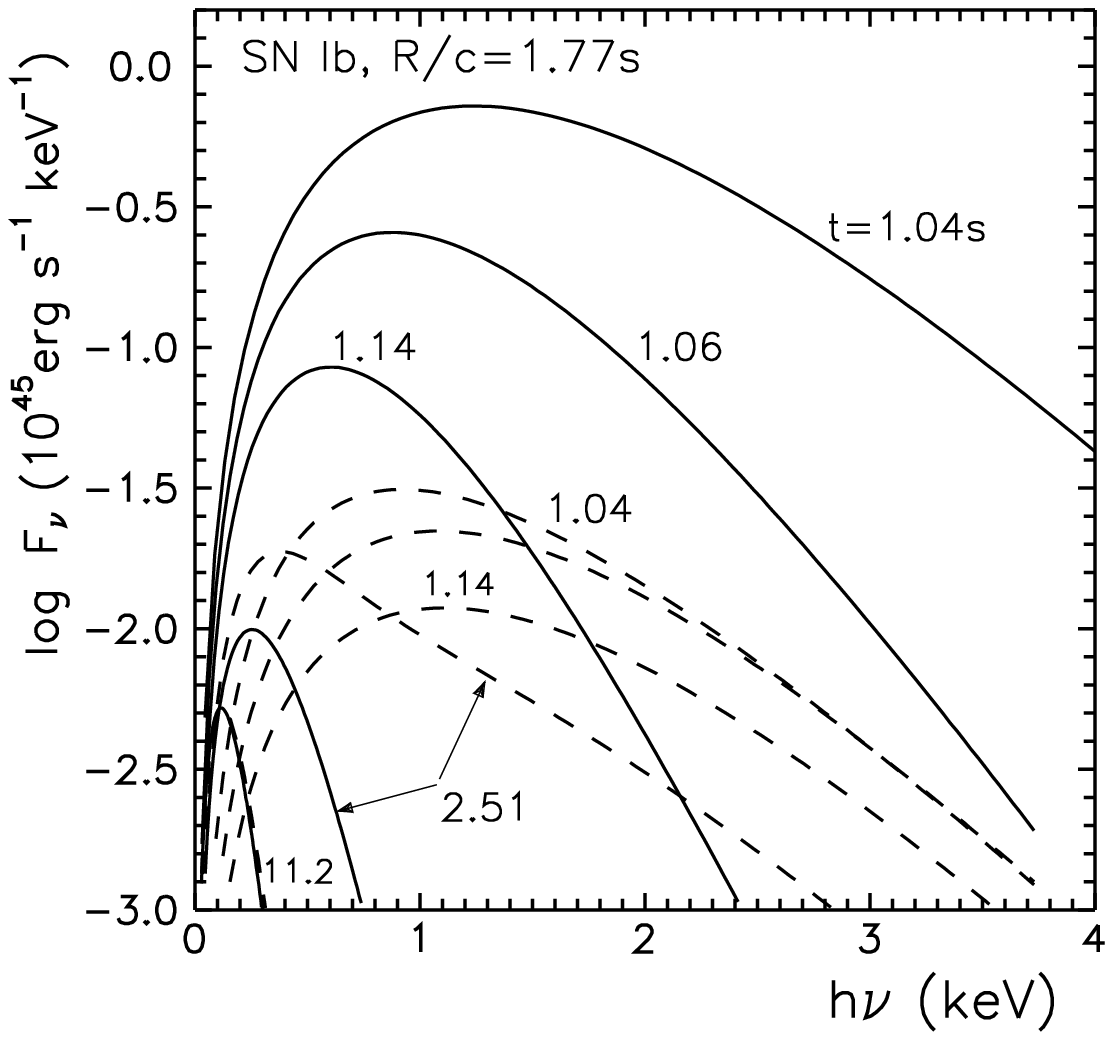}
\caption{Spectral luminosities for the model SN$\,$Ib
 (see \protect\fref{SNIb07} for the light curve).
\label{spIb}
}
\end{center}
\end{figure}

 In Figures %\ref{lc87a4LG},\ref{s1b7a2radafekoWacc4LG},\ref{s1b7a2SP}
 \ref{lc87a4LG}--\ref{s1b7a2SP}
 the results are presented calculated by {\sc stella} and {\sc rada} algorithms where
 radiation transfer is calculated numerically.
 As  in all the figures above the dashed
 lines correspond to the time-spread light curves while the solid
 ones show the case when the time-delay effect is neglected.
 The comparison with Figures %\ref{SNIb07},\ref{87AM18},\ref{spIb}
 \ref{SNIb07}--\ref{spIb}
 shows that exact calculation of
 radiation transfer retains the qualitative form of the the light curves and spectra
 and adds a number of details.

\begin{figure}
\begin{center}
\includegraphics[width=84mm]{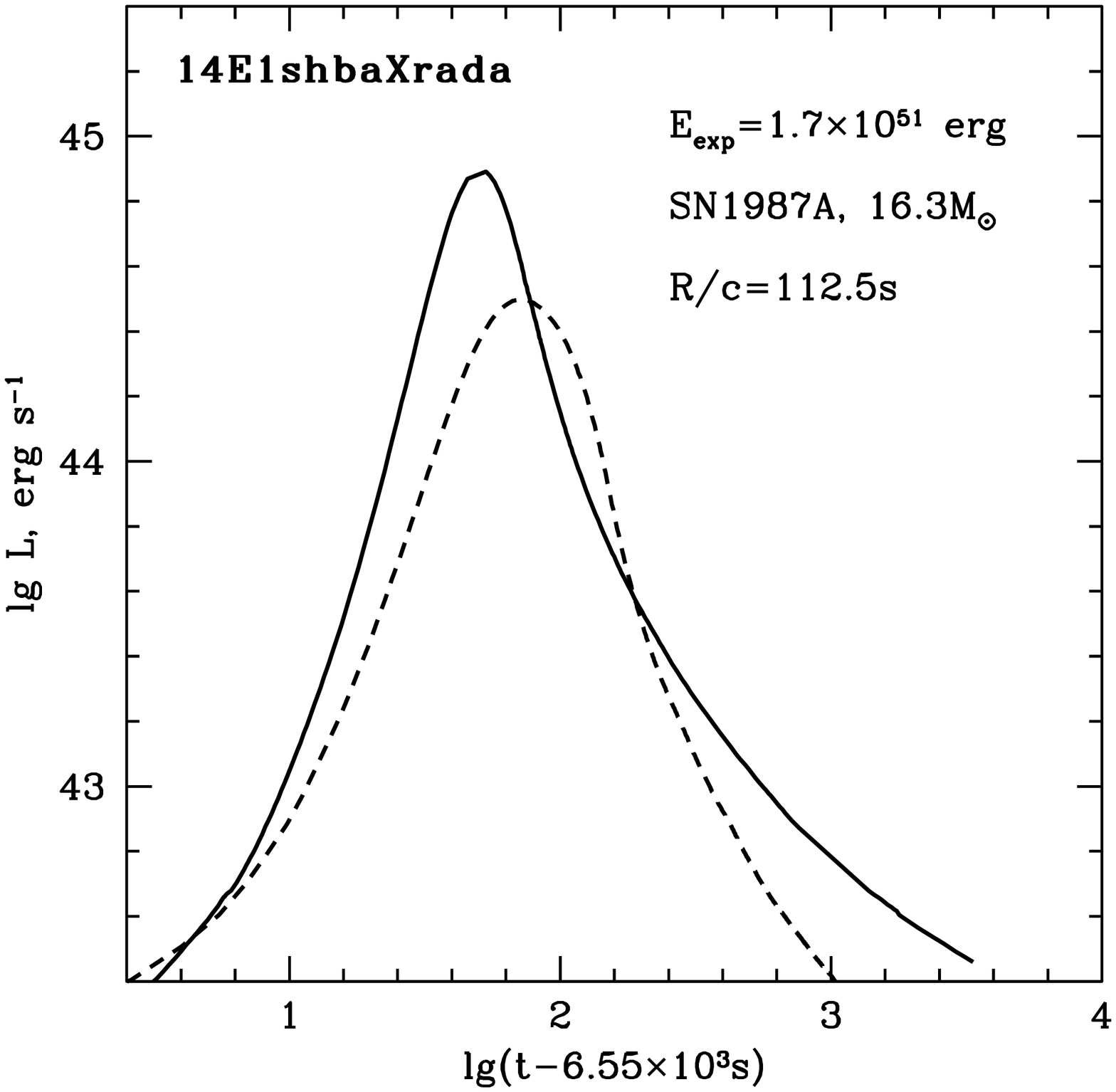}
\caption{The model SN$\,$1987A \NMS{16.3}
\label{lc87a4LG}
}
\end{center}
\end{figure}

\begin{figure}
\begin{center}
\includegraphics[width=84mm]{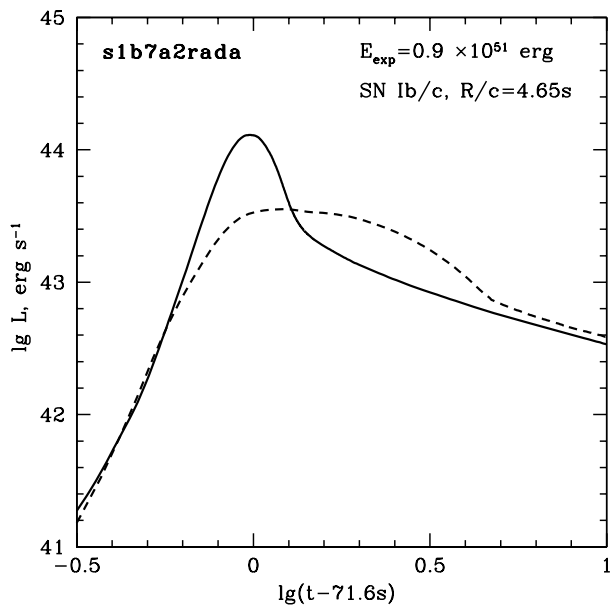}
\caption{The model SN$\,$Ib/c \NMS{3.2}
\label{s1b7a2radafekoWacc4LG}
}
\end{center}
\end{figure}

\begin{figure}
\begin{center}
\includegraphics[width=84mm]{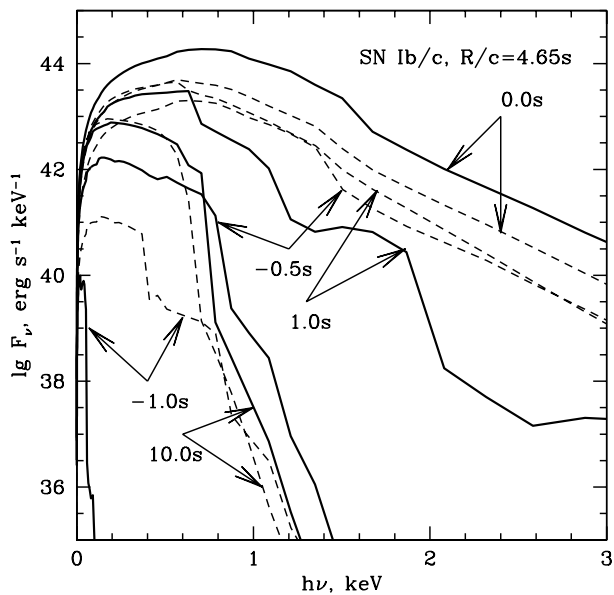}
\caption{Spectral luminosities for the model SN$\,$Ib/c
 (see \protect\fref{s1b7a2radafekoWacc4LG} for the light curve)
 at time stamps related to the maximum of the flux.
\label{s1b7a2SP}
}
\end{center}
\end{figure}

%__________________________________________________________________

\subsection{Shock breakout in the models of SN~1987A}

For our calculations of the SN~1987A outburst,
we used the presupernova model constructed by the
Tokyo group \citep{Shigeyama1987,ShigeyamaNomoto1990}.
For details of this model and the
variants of our runs, see \citet{Blinnikov1999,
Blinnikov2000}. The variants of this
series are designated as 14E1, 14E1.3 etc.,
where the number 14 corresponds to the ejecta mass in
solar masses \Msun (to be more precise, 14.7~\Msun).
The numbers after $E$ denote the explosion energy in units of $10^{51}$~erg.

Figures \ref{umr14e13} -- \ref{tpmbq2} show the velocity, density and temperature
profiles for one of these variants, 14E1.3.

\begin{figure}
{\includegraphics[width=84mm] {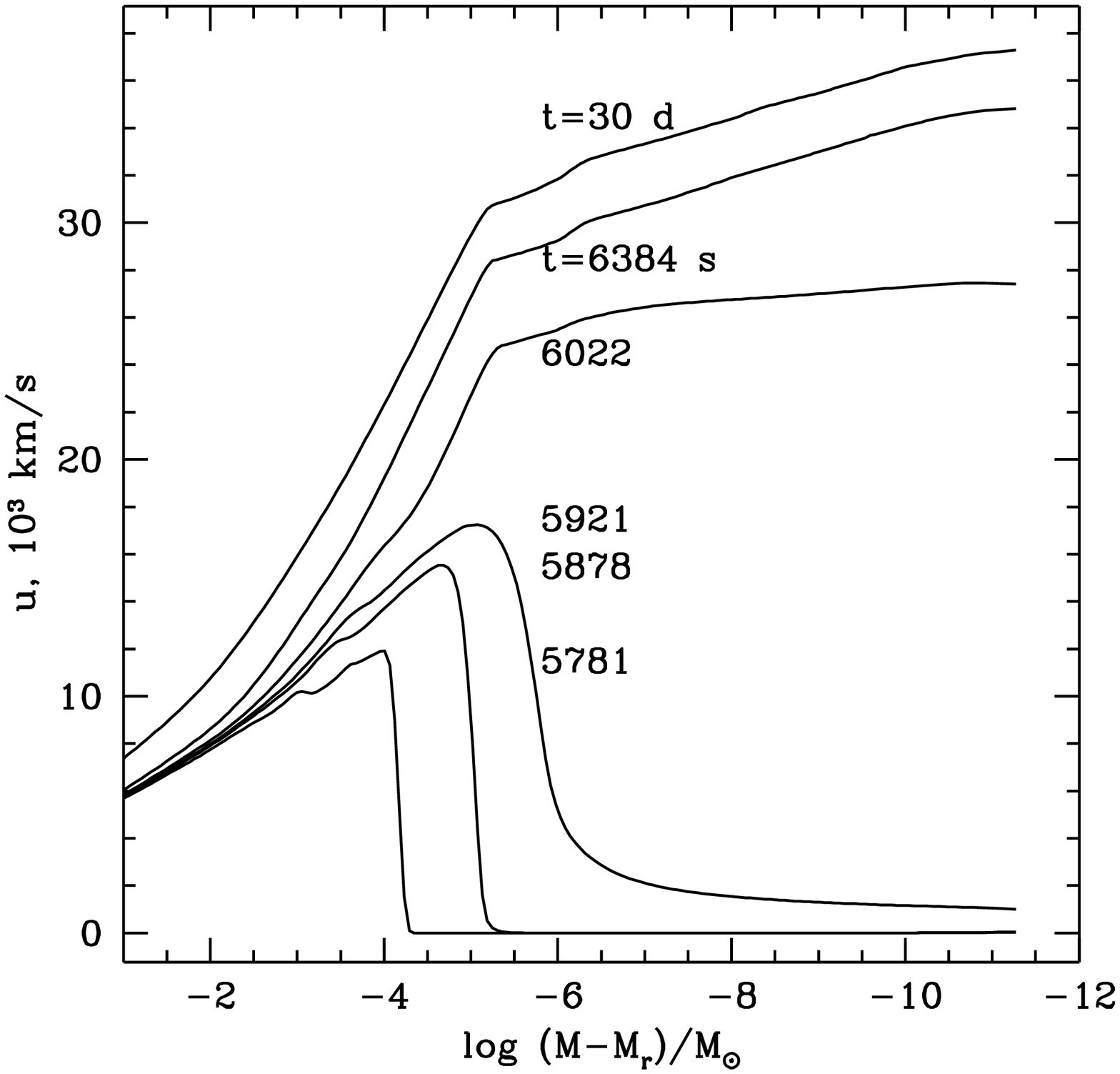}}
{\includegraphics[width=84mm] {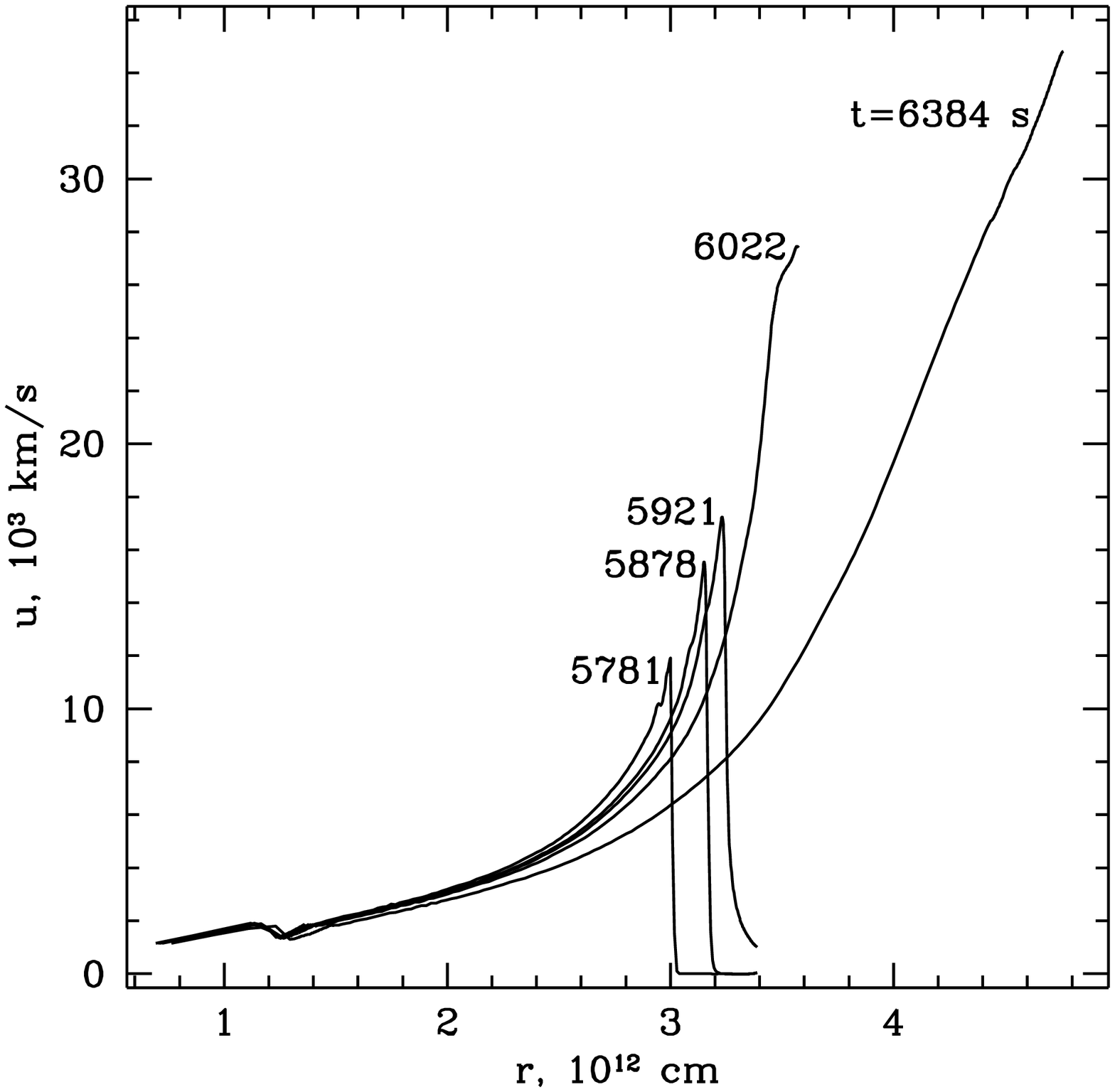}}
\caption{Matter velocity for the variant 14E1.3 at the epoch of shock breakout
         versus Lagrangean mass $M_r$ (top) and Eulerian
         radius $r$ (bottom) in the model for SN~1987A from \citet{Blinnikov1999}.
         The proper time is given near the curves.
\label{umr14e13}
}
\end{figure}

\begin{figure}
{\includegraphics[width=84mm] {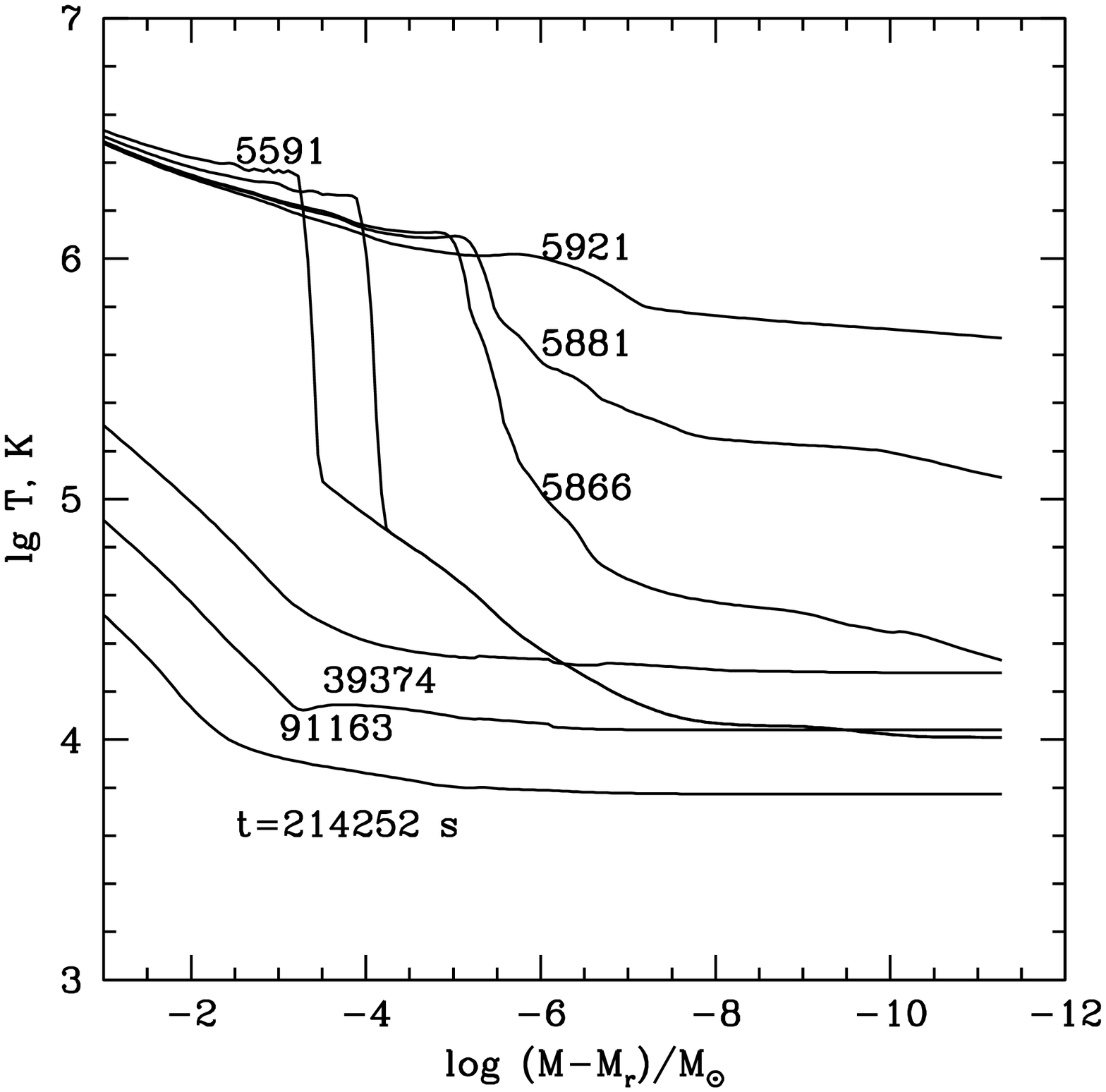}}
%\plotone{tpmbq2.ps}
{\includegraphics[width=84mm] {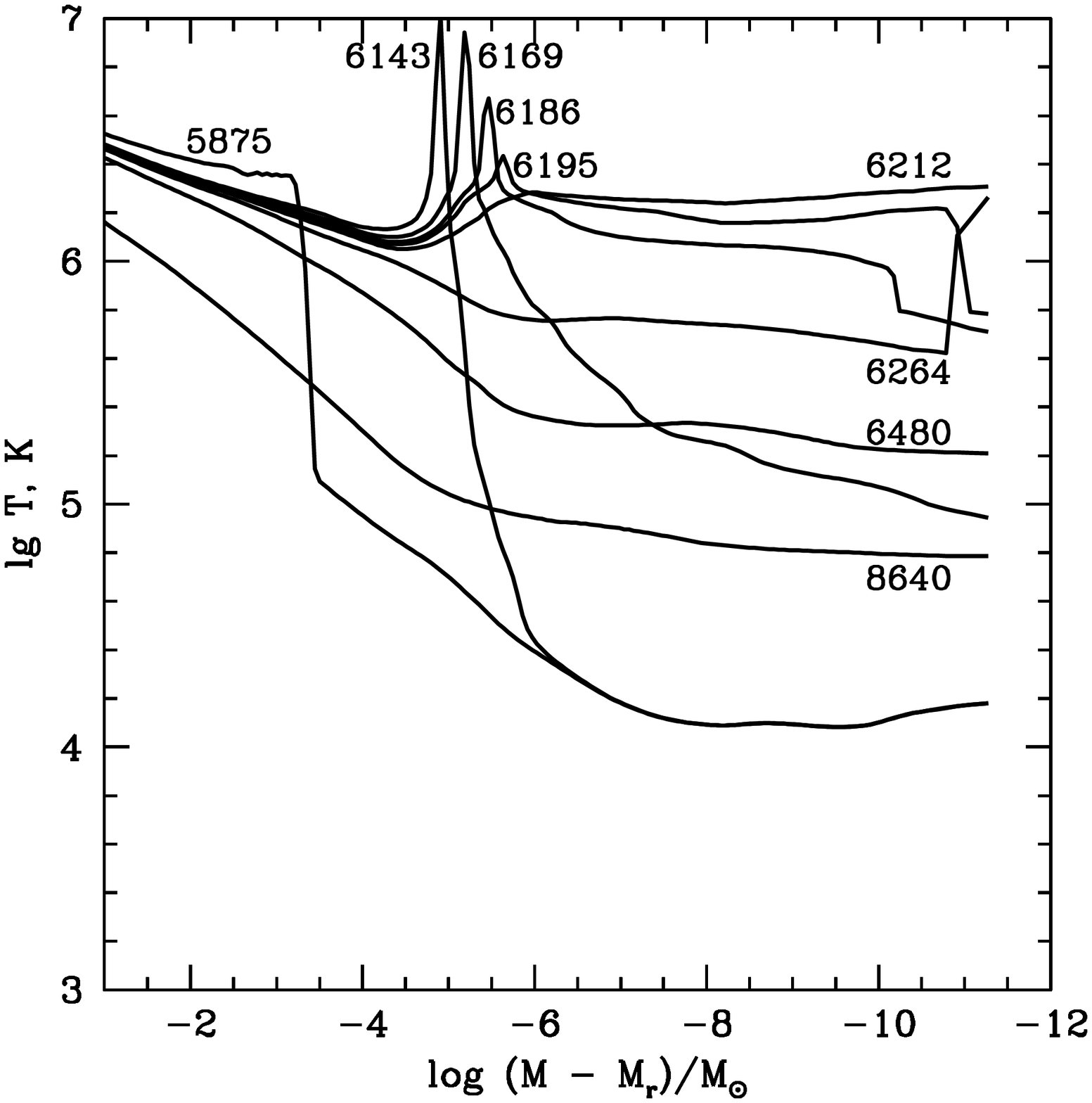}}
\caption{Matter temperature for the versions 14E1.3 (top) and 14E1X2 (bottom)
         at shock breakout versus Lagrangean mass $M_r$ measured from the surface.
         The time in seconds is given near the curves.
\label{tpmbq2}
}
\end{figure}

In the outermost layers, condition (\ref{shf}) is violated,
the losses through radiation become significant, and
the shock acceleration predicted by the self-similar
solution ceases. This behavior is excellently seen from Fig.~\ref{vm14e1tau},
where the velocity profiles are shown as a
function of Lagrangean mass and optical depth $\tau$ in the model 14E1 for SN~1987A \citep{Blinnikov1999}.
The ``breakout'' of photons from behind the shock front
occurs precisely at $\tau\sim10$.
These photons slightly
accelerate the overlying layers of matter, but the cumulation
of energy at a low mass is already inefficient
due to great losses through radiation. The asymptotic
ejecta velocity distribution in mass for this variant is
given in the table from \citet{Blinnikov1999}. % ~ \ Ref (veltab).

\begin{figure}
{\includegraphics[width=84mm] {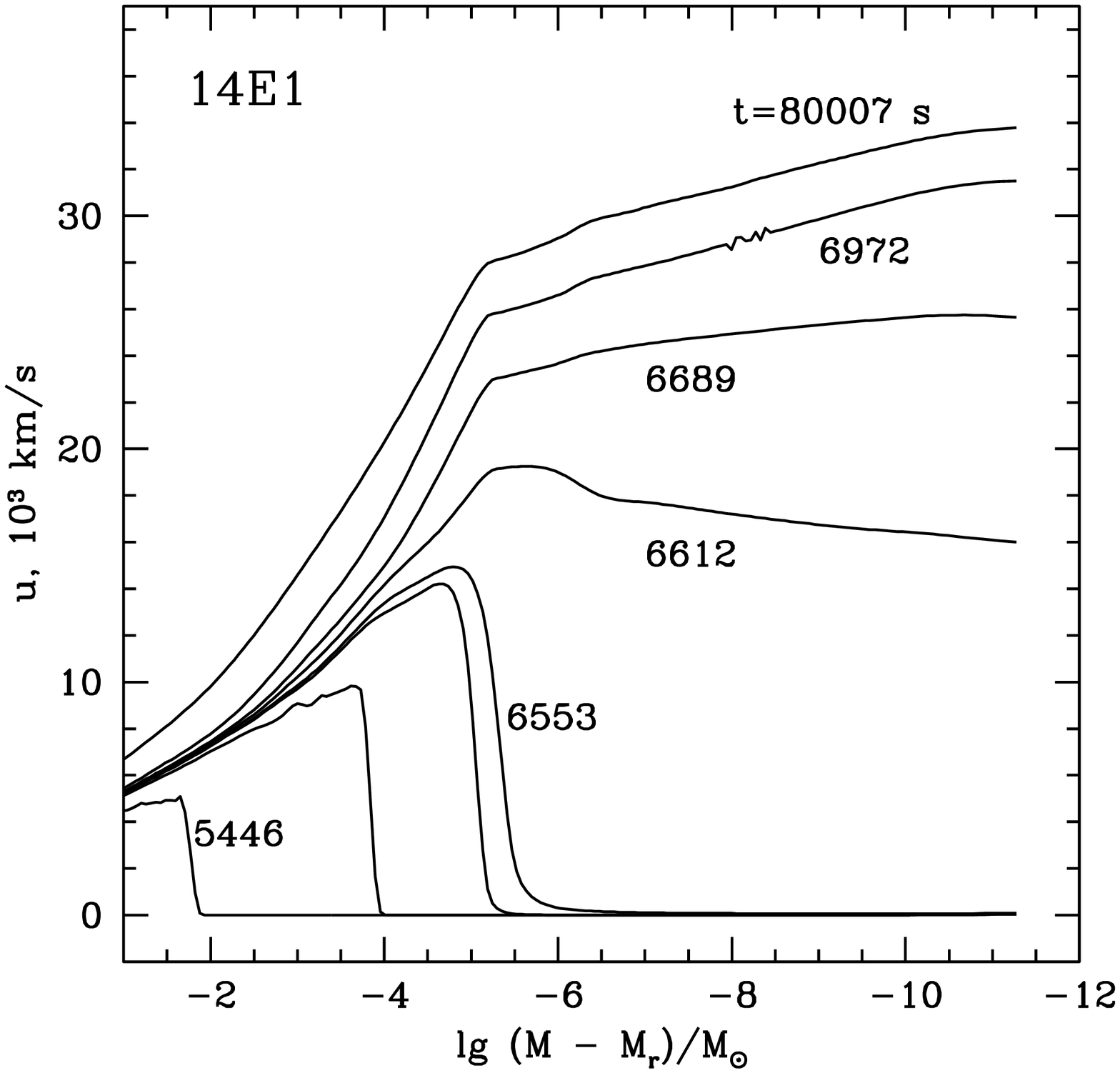}}
{\includegraphics[width=84mm] {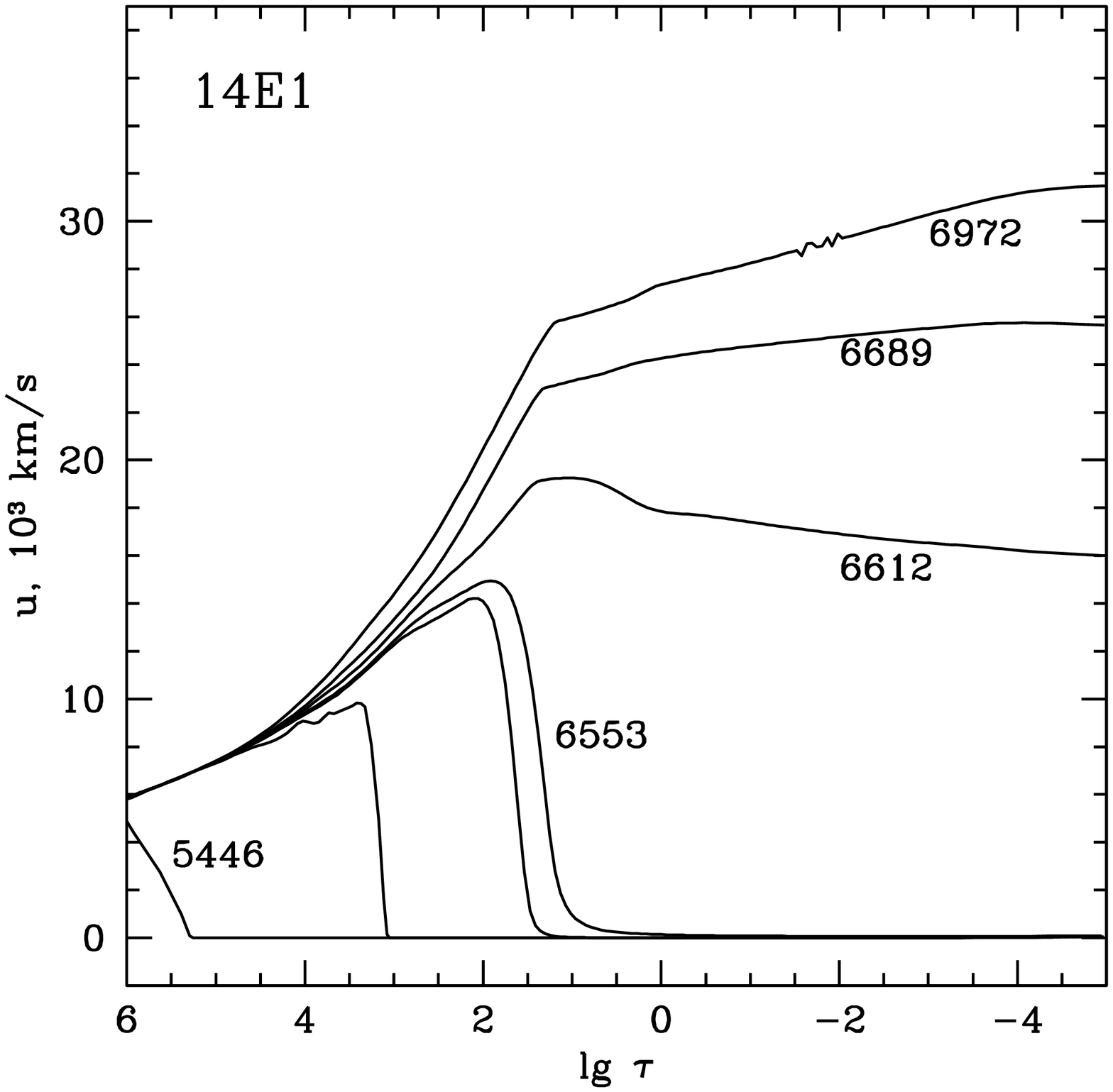}}
\caption{Velocity profiles as a function of Lagrangean mass (top) and
         optical depth $\tau$ (bottom) in the model 14E1 for SN~1987A from
         \citet{Blinnikov2000}.
         The time in seconds is given near the curves.
\label{vm14e1tau}
}
\end{figure}

The velocity distribution of the ejected mass, the
mass spectrum, is of great importance in astrophysical
applications. The mass spectrum $M(u)$ with a
velocity greater than $u$ can be predicted analytically at
the self-similar stage. Analysis shows that the mass
spectrum for the outer part of the envelope is defined
by a simple relation. It obeys the Nadyozhin--Frank-Kamenetskii law
\begin{equation}
 M(u)= \mathrm{const}\cdot u^{-\frac{n+1}{\lambda}} \sim u^{-7.2} \; ,
\label{NadFK}
\end{equation}
where $M(u)$ is the mass with a velocity greater
than $u$, the numerical value of the exponent is written
out for the case of interest to us \citep[see][]{NadezhinFrankKamenetskii1964translation},
$n$ is the exponent
in the law $\rho \propto (R-r)^n$ \citep{Grasberg1981}, and
$\lambda$ is the eigenvalue of the self-similar problem
\citep{GandelmanFrankKamenetskii1956,Sakurai1960}.

The mass of the outer part of the envelope described
by law (\ref{NadFK}) is determined by the applicability
conditions for self-similar solutions.
For example, our calculation gives $u=5.5\times 10^3$ km/s at $M=0.1 M_\odot$.
The prediction of $u=30\times 10^3$km/s must be at $M\approx 2\times 10^{-6}M_\odot$.

Figure \ref{umr14e13} shows similar velocity profiles for the
variant 14E1.3, where the explosion energy is higher
than that in 14E1 by 30\%.
The picture is very similar
to Fig.$\,$\ref{vm14e1tau}, but now the breakout time
decreased and the asymptotic ejecta velocity slightly
increased.
It is interesting to see how this picture
appears in Eulerian coordinates
(Fig. \ref{umr14e13}).
Figure \ref{rhomrbq0} shows how the density peak is formed in the outer layers
due to inefficient gas acceleration. Our numerical
experiments show that the development of one, two, or more peaks is possible:
this is analogous of the loss of stability of sequential
harmonics. In this case, we see them
at the nonlinear stage. How many peaks there are and
how strong they are depends on the shock strength,
on the density profile, on the absorption and emission
coefficients, and (in our calculations) on the grid fragmentation
and artificial dissipation parameters. This
question could be investigated, but such a problem
would be purely academic in the one-dimensional
approximation --- because even the formation of the
first peak must lead to the development of multidimensional
instabilities in it (in a thin layer) and to
layer fragmentation. Another analogy is a steepening
of a great wave coming on shore; it can be seen how
increasingly high harmonics are formed from one sine
wave followed by breaking --- everything flies away in
small splashes (multidimensionality).

\begin{figure}
{\includegraphics[width=84mm] {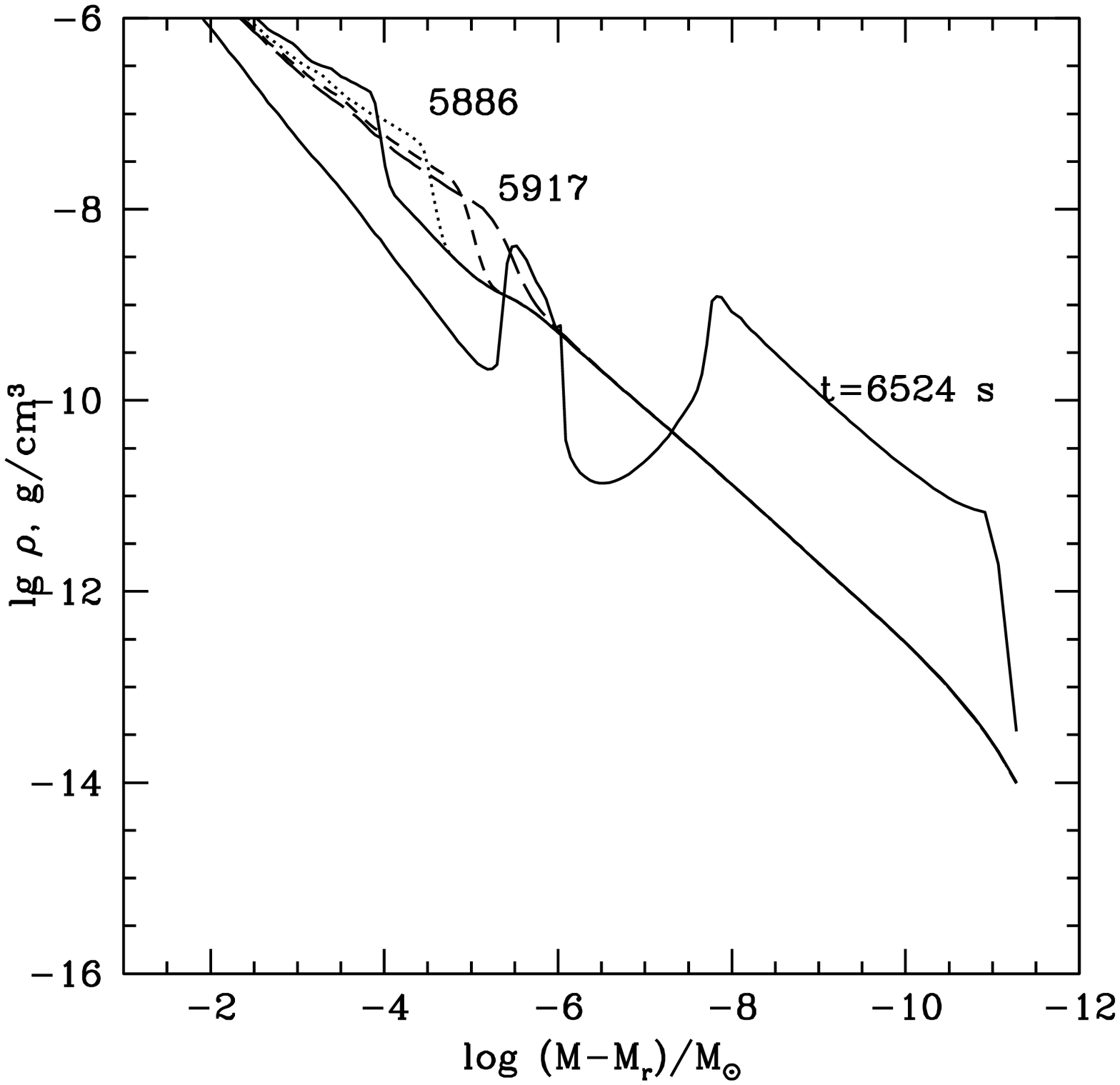}}
{\includegraphics[width=84mm] {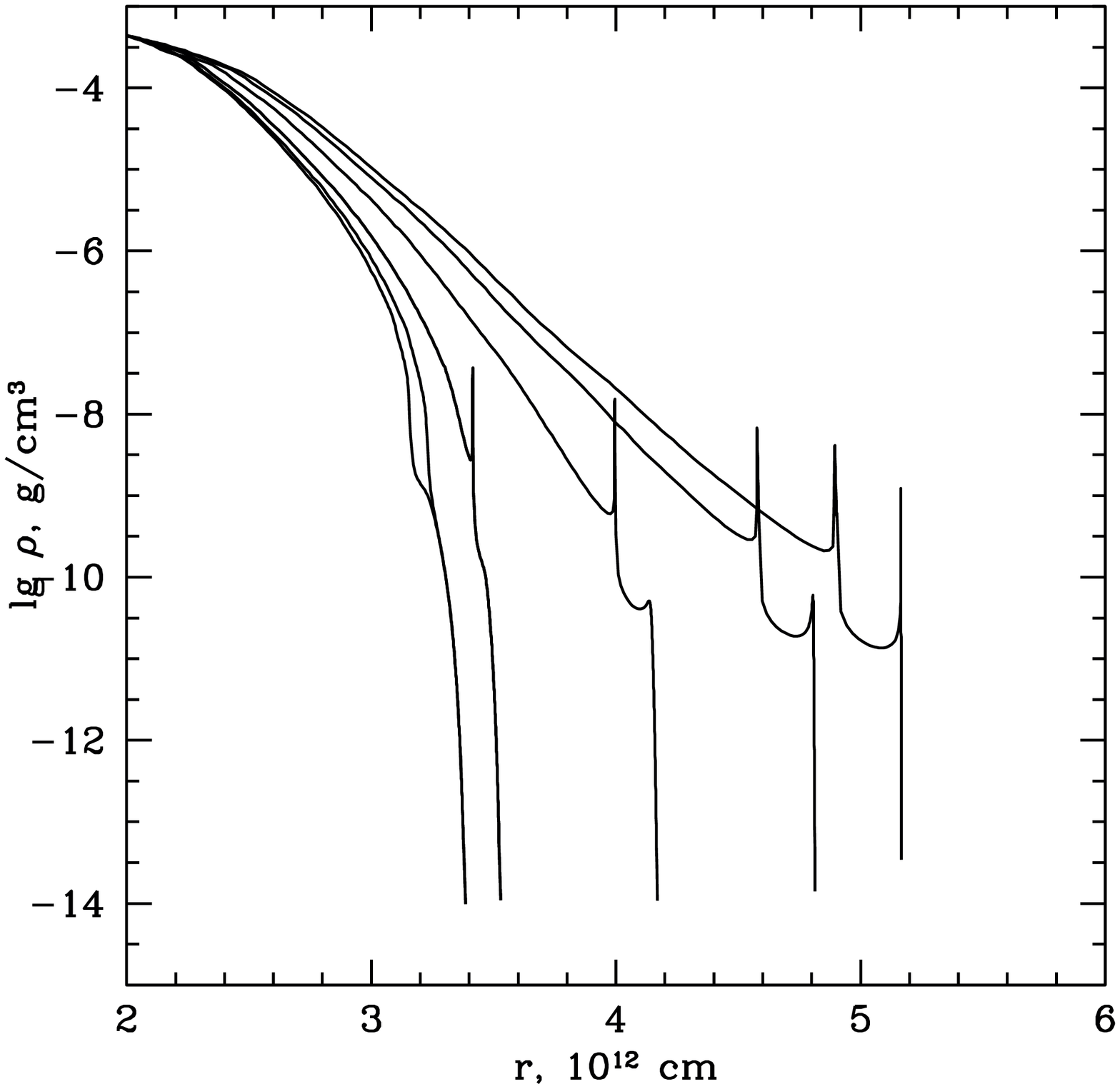}}
\caption{Matter density for the variant 14E1.3 at the epoch of shock breakout
         versus Lagrangean mass $M_r$ (top)
         and Eulerian radius $r$ (bottom).
         The proper time is given near the curves.
\label{rhomrbq0}
}
\end{figure}

Let us now consider the influence of opacity on
the parameters of the shock breaking out from a
presupernova. We see from
\rref{tpmbq2} that the matter
temperature behind the shock front $T$ does not exceed
$\sim3\cdot10^6$ K.
\citet{EnsmanBurrows1992} obtained $T$ higher by almost two orders of magnitude, while the
color temperature of the radiation agrees satisfactorily
with our value in \rref{14e1lumb}.
What is the cause of
this contradiction? The point is that the description
of opacity in \citet{EnsmanBurrows1992} was too
crude. These authors simply subtracted the Thomson
opacity from the Rosseland mean $\chi_{\rm R}$ and assumed
this quantity to be equal to the Planck mean for true
absorption $\alpha_{\rm P}$.
Since the Thomson scattering at high
temperatures dominates in
$\chi_{\rm R}$ and is higher than absorption
by a huge factor, the subtraction of the two close
numbers leads to very large errors. In addition, the
identification of the Rosseland and Planck means is
an improper procedure \citep{ZeldovichRaizer1966}.
In our calculations, the true absorption is calculated
separately from the total extinction, i.e., always fairly
accurately, and no problems with the averaging of
opacity over the entire spectrum arise in multigroup
calculations.
\citet{EnsmanBurrows1992} explain
the high matter temperature $T$
by the formation of a
viscous jump in transparent layers with strong gas
heating (as was discussed above, there is no viscous
jump in deep layers in a strong, radiation-dominated
shock).

\begin{figure}
\includegraphics[width=84mm] {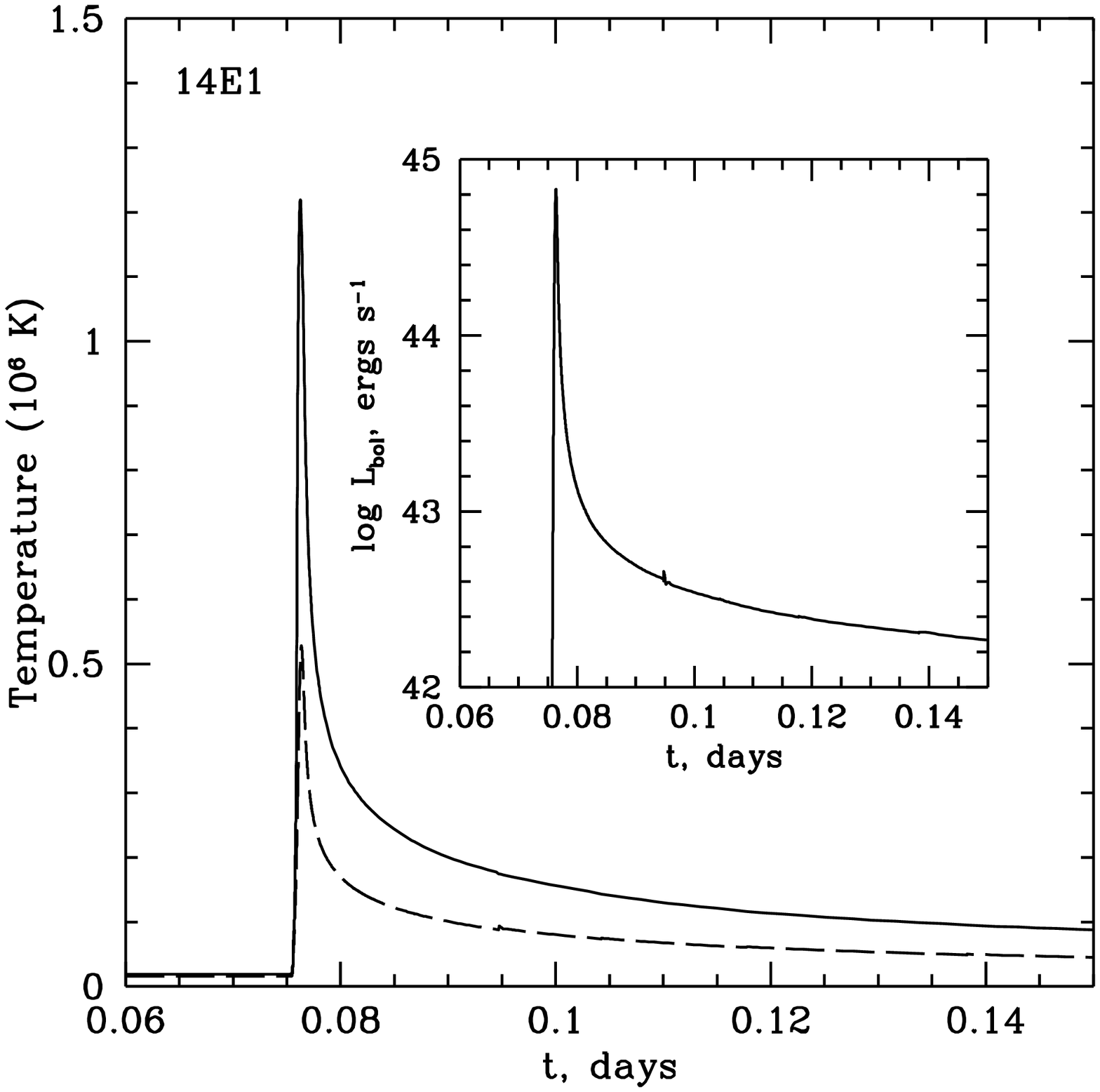}
\includegraphics[width=84mm] {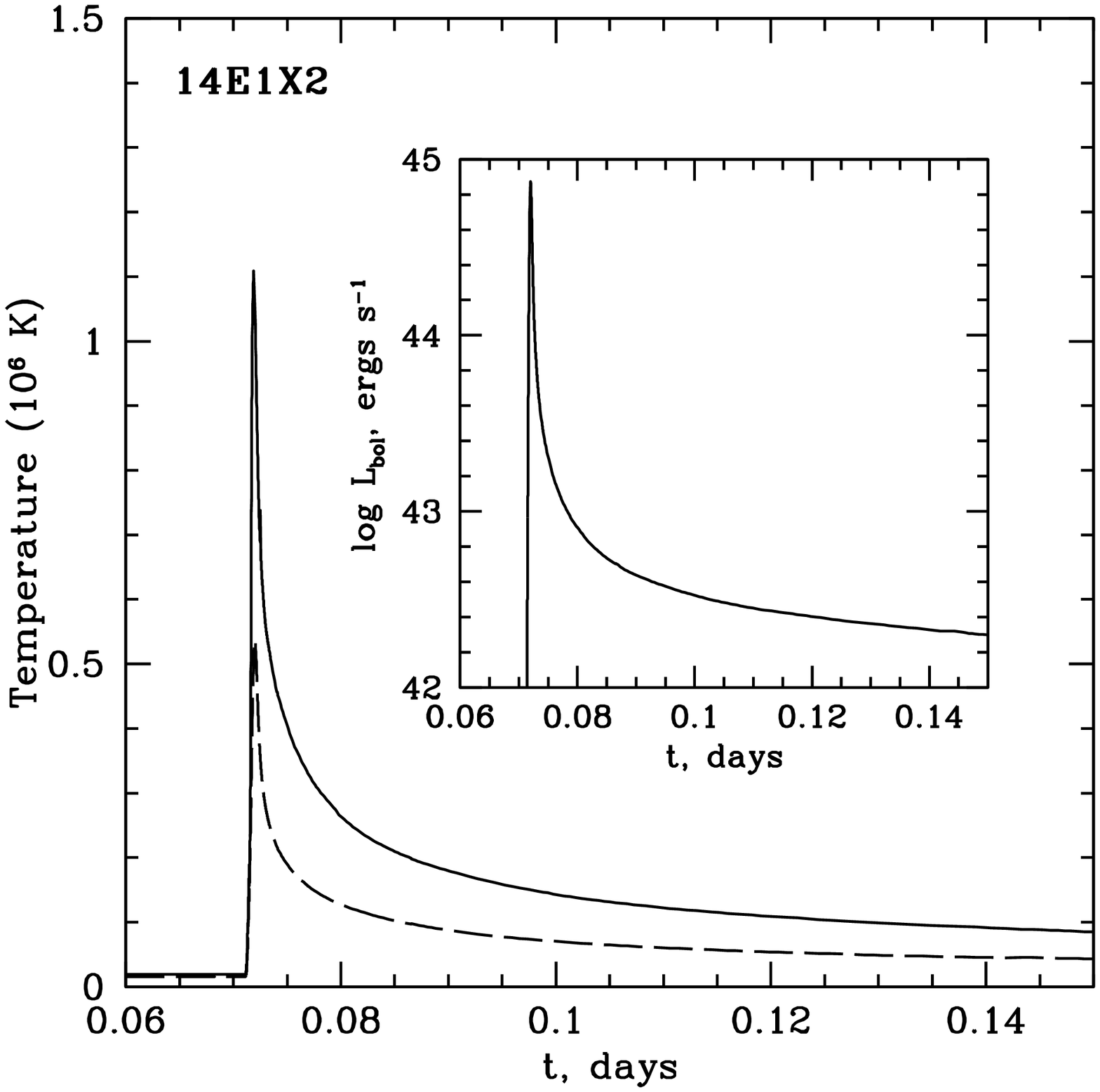}
\caption{Bolometric light curve (in the inset) and color (solid line) and effective
         (dashed line) temperatures for two versions of the model 14E1 (see the text).
\label{14e1lumb}
}
\end{figure}

To check the influence of opacity on T in a shock,
we carried out a numerical experiment by replacing
the complete opacity calculation algorithm with the
following approximate procedure. The code that we
developed for the equation of state solves the Saha
system of equations for an arbitrary number of elements
at all ionization stages. In the calculations that
we will now describe, the ionization is described in
the ``mean-ion'' approximation by \citet{Raizer1959a} (see also the book by
\citet{ZeldovichRaizer1966}).

We obtain the continuum opacity $\chi_{\rm c}$
needed for
our calculations using the following procedure (only
in this numerical experiment!).
Initially, we take an
expression for the monochromatic absorption coefficient
similar to that used by \citet{Vitense1951} and based
on the Kramers approximation for our mixture of elements.
This expression, plus the Thomson opacity,
is not used directly as $\chi_{\rm c}$ but is used only to obtain
the Rosseland mean $\chi_{\rm R}$ through approximate integration
proposed by \citep{Raizer1959b,ZeldovichRaizer1966}. We obtain the true absorption
coefficient $\alpha_{\rm P}$ just as was done by
\citet{EnsmanBurrows1992}, thereby qualitatively reproducing the
algorithm from their paper, and use $\alpha_{\rm P}$ instead of
the correct monochromatic absorption coefficient in
our multigroup calculations. We will call the opacity
obtained in this way Zeldovich--Raizer one and will
denote it by ZR.

\begin{figure}
{\includegraphics[width=84mm] {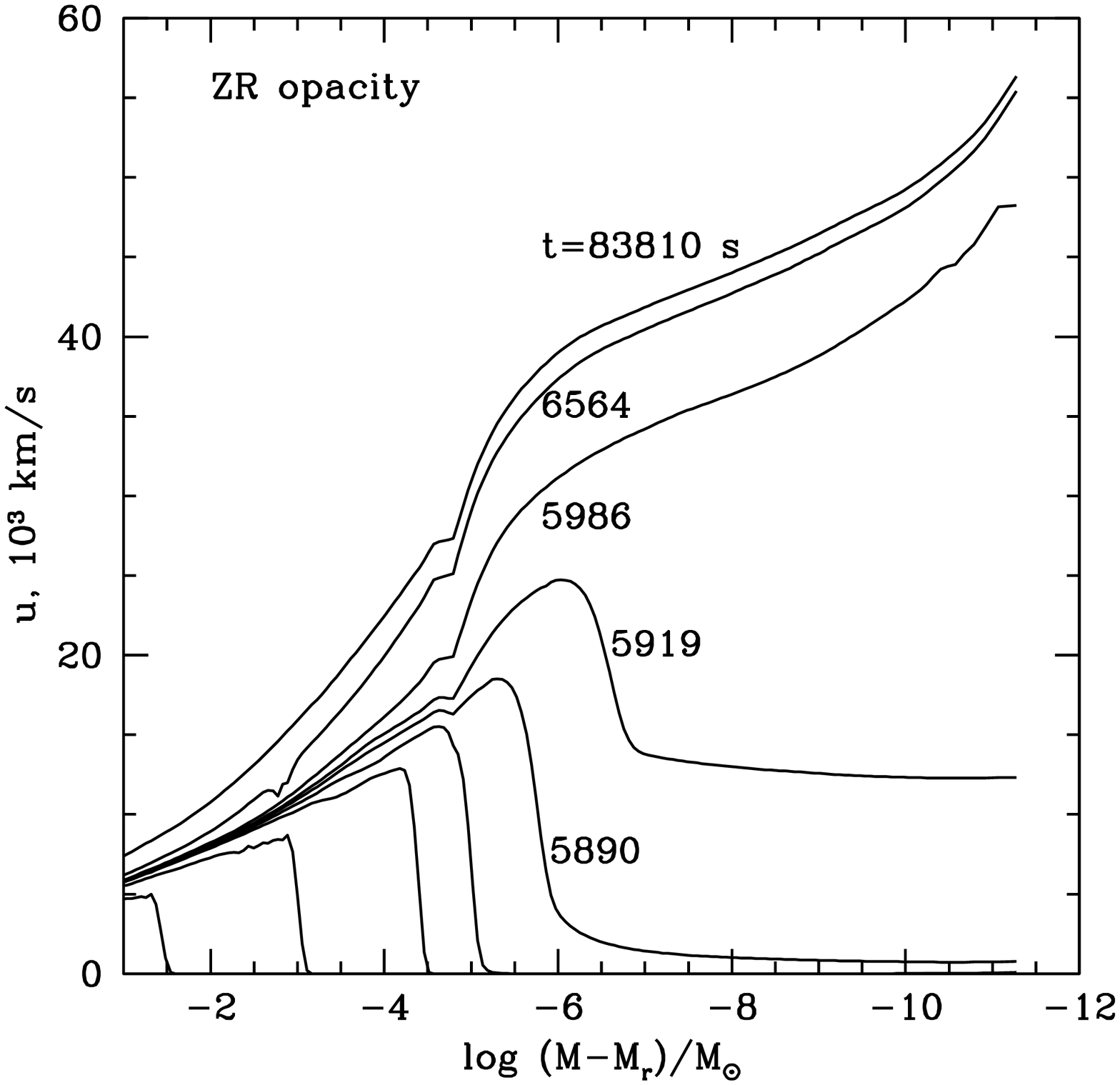}}
{\includegraphics[width=84mm] {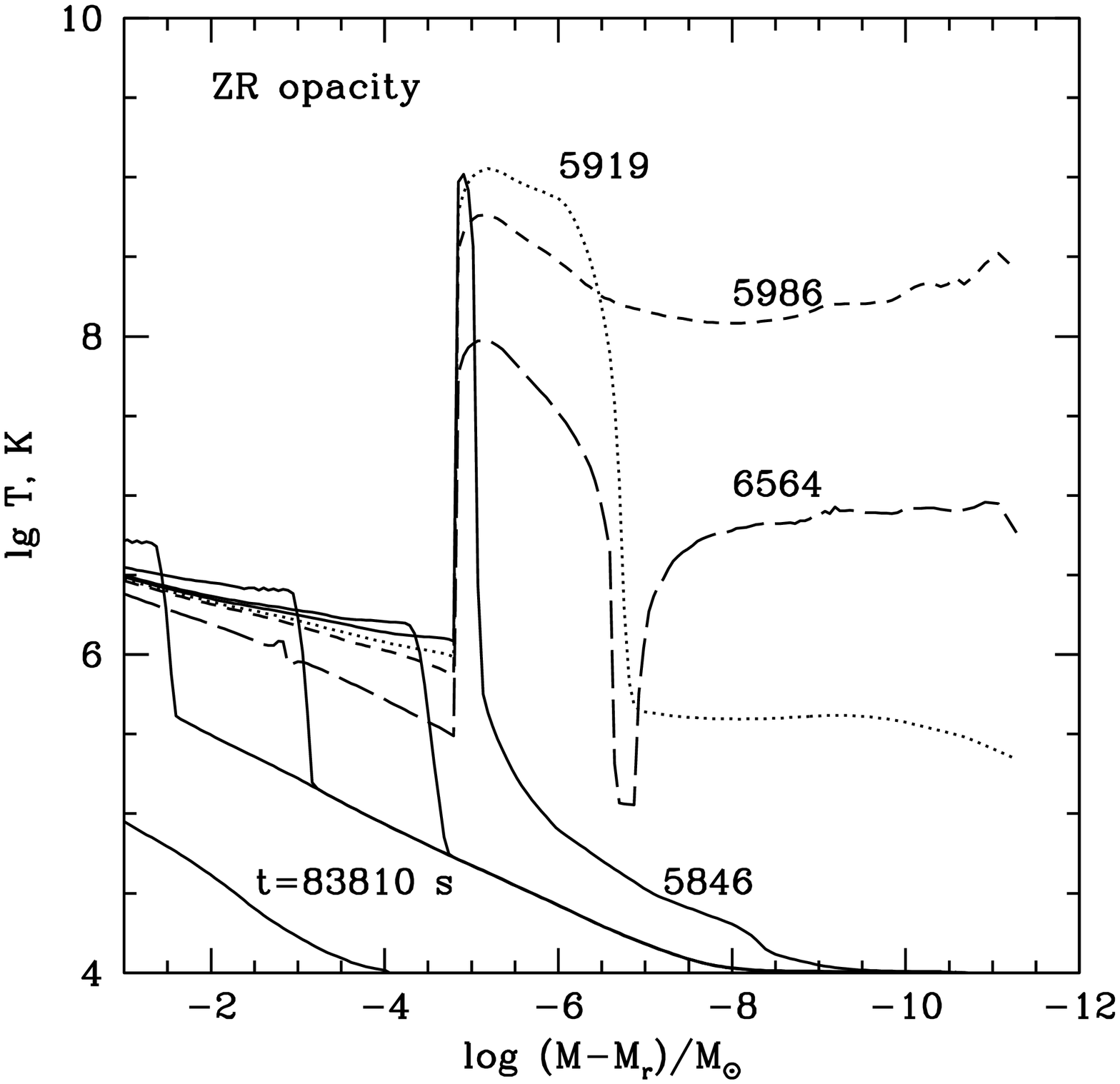}}
\caption{Matter velocity ($10^3$ km/s, top) and temperature (bottom) for the variant 14E1.3
         with Zeldovich--Raizer (ZR) opacity at the
         epoch of shock breakout versus Lagrangean mass $M_r$.
         The proper time is given near the curves.
\label{vmhc}
}
\end{figure}

The matter velocity and temperature for the variant 14E1.3
with ZR opacity at the epoch of shock
breakout are shown in \rref{vmhc}. %, \ Ref (tpmhc).
We immediately
see that the matter in this variant speeds up to
higher velocities and heats up in the viscous jump
to enormous temperatures,
$T\sim 10^9$ K.
Both these
effects are explained by underestimated true absorption:
the quasi-adiabatic acceleration of the matter
in a self-similar shock
\citep{GandelmanFrankKamenetskii1956,Sakurai1960}
continues longer (cf. the velocity profile in \rref{umr14e13}),
while the matter
behind the density jump cannot efficiently give up
heat and heats up virtually to values corresponding
to an adiabatic shock.

Figure ~\ref{blc87a} compares the calculated bolometric light
curves for model 14E1shbaXradaNhm23 of the SN 1987A presupernova
\citep{Blinnikov2000}.
%In contrast to the
%model of a type Ib/c presupernova considered above,
%the velocity of the outer layers here is lower by an order
%of magnitude and
The contribution from the {\sc rada}
algorithm to the computation manifests itself mainly
in a careful allowance for the radiation time delay, which
leads to a decrease in radiation flux at maximum and
to a broadening of the light curve peak.

It should be noted that light curve calculated by {\sc rada}
algorithm does not have second peak presented earlier \citep{Tolstov2010}
due to more accurate allowance for boundary conditions.

\begin{figure}
\centering
\includegraphics[width=84mm]{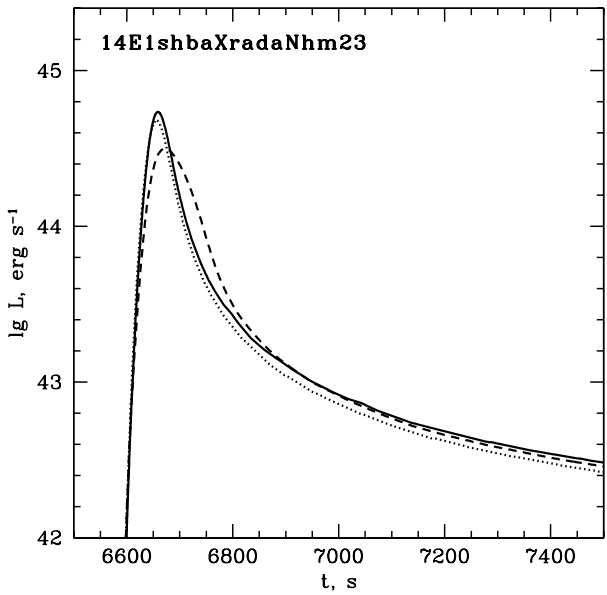}
\caption{Comparison of the bolometric light curves at the epoch
         of shock breakout  for an
         SN1987A-type presupernova model. The solid and dotted lines represent,
         respectively, the {\sc stella} and {\sc rada} calculations
         in the comoving frame of reference.
         %which do not take into account radiation time delay.
         Dashed line represents {\sc rada} calculations in observer's frame of reference
         taking into account radiation time delay.
}
\label{blc87a}
\end{figure}

  \subsection{The dependence of the parameters of the flash from the explosion energy and the radius of presupernova}

\begin{table*}
\begin{minipage}{126mm}
\caption{Predictions for the first brightness peak
\label{earlight}}
\begin{center}
\begin{tabular}{llllllll}
\hline
\hline
 \\[-3mm]
variant & $t$, d & $L_{\rm bol}$, erg/s  & $T_{\rm c}$, K & $T_{\rm eff}$, K & $R_{\tau=2/3}$, cm & $\int_0^{2 {\rm d}}Ldt$, erg\\
%     & $T_{\rm eff}$, K & $R_{\tau=2/3}$, cm & $\int_0^{2 {\rm d}}Ldt$, erg\\
\hline
 \\[-3mm]
14E0.7     & .08960 & 4.217\e{44} & 1.074\e{6} & 4.71\e{5} & 3.32\e{12}& 1.07\e{47}\\
14E1       & .07637 & 6.751\e{44} & 1.219\e{6} & 5.28\e{5} & 3.32\e{12}& 1.40\e{47}\\
14E1.3     & .06726 & 9.466\e{44} & 1.339\e{6} & 5.73\e{5} & 3.32\e{12}& 1.77\e{47}\\
14E1.34R   & .05620 & 9.175\e{44} & 1.451\e{6} & 6.26\e{5} & 2.74\e{12}& 1.36\e{47}\\
14E1.4R6   & .07520 & 1.013\e{45} & 1.267\e{6} & 5.30\e{5} & 3.95\e{12}& 2.38\e{47}\\
\hline
% \\[-3mm]
%variant %& $t$, d & $L_{\rm bol}$, erg/s  & $T_{\rm c}$, K
%
%\hline
% \\[-3mm]
%14E0.7     & 4.71\e{5} & 3.32\e{12}& 1.07\e{47}\\
%14E1       & 5.28\e{5} & 3.32\e{12}& 1.40\e{47}\\
%14E1.3     & 5.73\e{5} & 3.32\e{12}& 1.77\e{47}\\
%14E1.34R   & 6.26\e{5} & 2.74\e{12}& 1.36\e{47}\\
%14E1.4R6   & 5.30\e{5} & 3.95\e{12}& 2.38\e{47}\\
\hline
\end{tabular}
\end{center}
Note. The $T_{\rm eff}$ maximum almost coincides in time with the
$L_{\rm bol}$ peak, while the $T_{\rm c}$ maximum occurs $\sim 100$ s earlier.
\end{minipage}
\end{table*}

In the models for SN~1987A, the explosion energy
was varied at constant presupernova radius. In addition,
we constructed two more models with greatly
differing radii, 10 and 300 \Rsun.
The presupernova
model 14E1 for SN~1987A
with initial radius $R_0=48$~\Rsun, ejection mass $M=14.7$~\Msun,
and $E_{\rm kin}=1.0 \cdot 10^{51}$ erg was taken as the initial one.
The explosion
energy in these models was the same. We constructed
models with radii slightly different from the standard
value: decreased $R_0=40$~\Rsun (e.g., 14E1.34R) and increased $R_0=58$~\Rsun,
model 14E1.4R6.
In addition, we
also constructed models with a great increase in the
radius to 300 \Rsun and with its decrease to
10 \Rsun.
The
density distribution in the new models was obtained
homologically from the model 14E1.
The outburst
parameters are presented in Table. ~ \ref{earlight} and \ref{velasym}.

The kinetic energies of the ejecta with radii of 10 and 300 \Rsun\ slightly differ, because to achieve complete
coincidence between their values, the calculation
should be performed several times by iterating
explosions with trial energy release at the center.
We see from \rref{earltef3} that the outburst time naturally
increases with radius (for a discussion of this dependence, Blinnikov et al, 2000),% (\ ref (tprop)),
while the peak
luminosity depends weakly on radius: when the radius
increased by a factor of 30, from 10 to 300 \Rsun\, the
luminosity peak decreased approximately by a factor of 2.
The effective temperature maxima for three
versions of the model 14E1
with different radii shown in \rref{earltef3} depend more strongly on $R_0$:
in accordance
with the definition of $L=4 \pi \sigma T_{\rm eff}^4R_0^2$,
we have $T_{\rm eff} \propto R_0^{-1/2}$ at constant $L$.
Recall that the color temperature
is approximately a factor of 3 higher than the
effective one in these outbursts, as can be seen from table~\ref{earlight}.

\begin{figure}
\includegraphics[width=84mm] {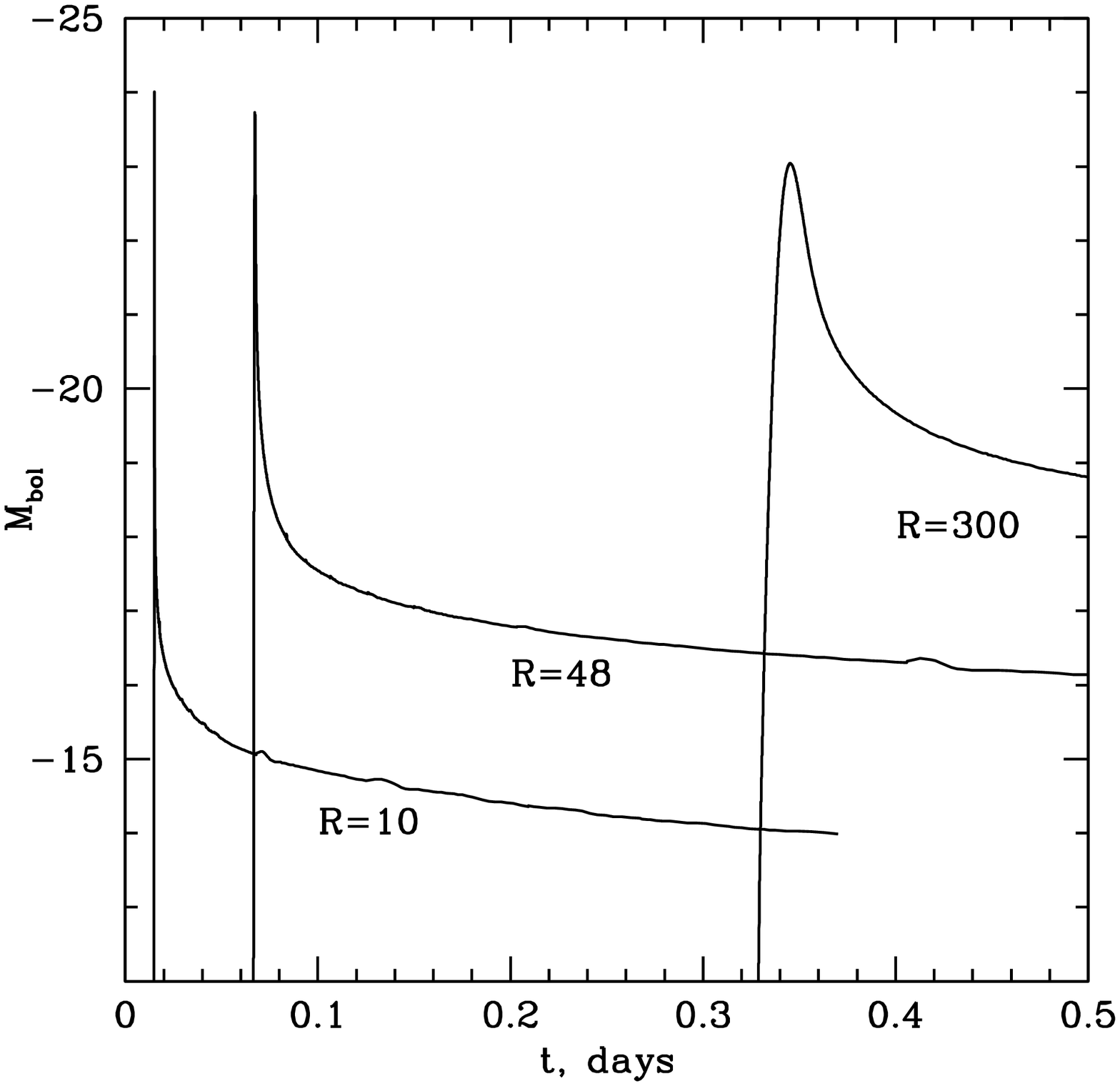}
\includegraphics[width=84mm] {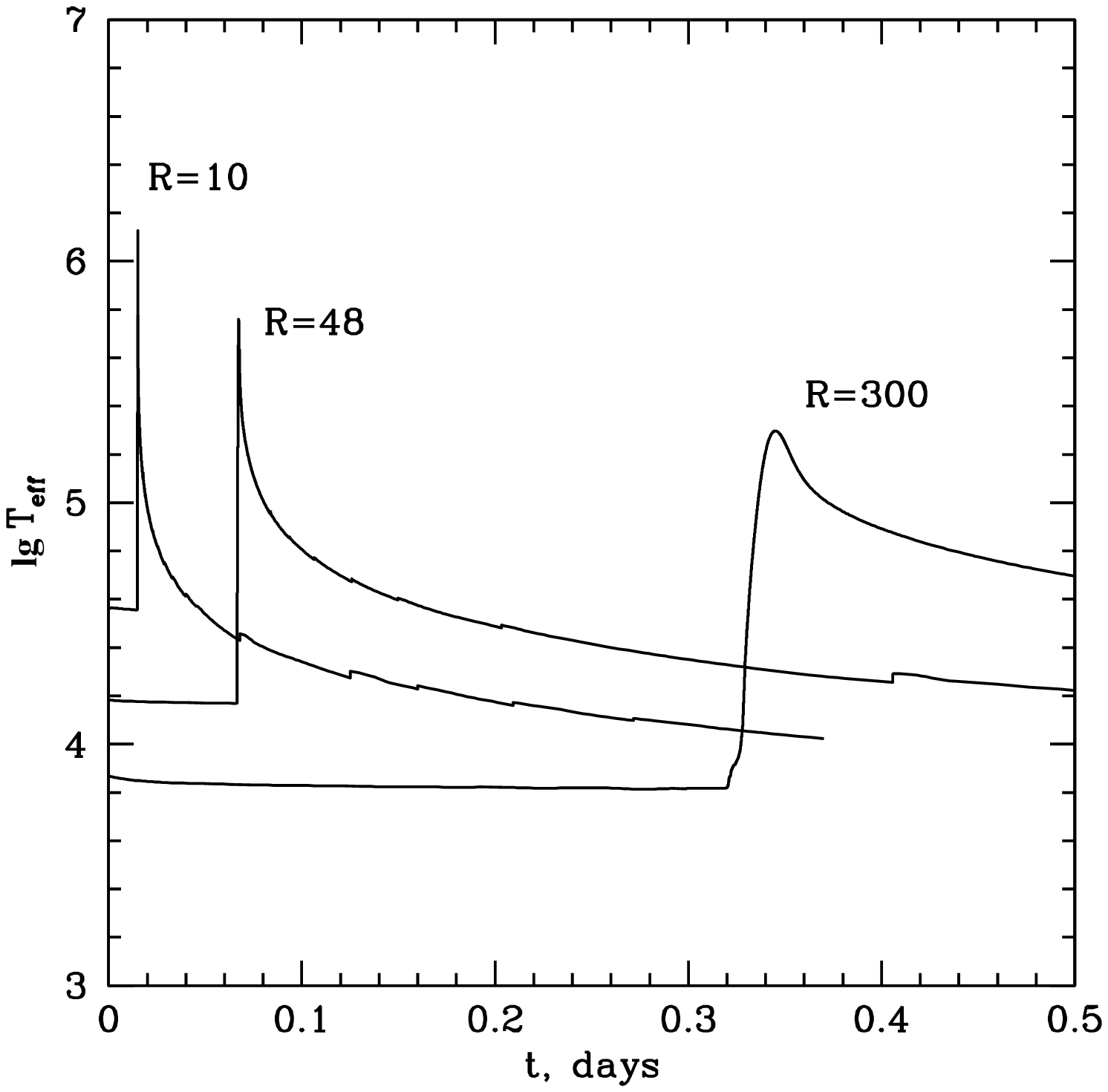}
\caption{Absolute bolometric magnitudes (top) and effective temperature
        (bottom) for three versions of the model 14E1
         with different radii at the epoch of shock breakout.
         The radii are given in solar units.
\label{earltef3}
}
\end{figure}

An important parameter of a supernova is the
maximum ejecta velocity. For various versions of the
series 14E this parameter is given in
Table.~\ref{velasym}.
Table.~\ref{velopa} shows the influence of various assumptions about
opacity on this parameter. Standard opacity suggests
allowance for spectral lines in the approximation of a
static medium (i.e., the values that correspond to the
pre-shock region are taken). ``High'' opacity is the
case where the spectral lines are everywhere broadened
by the maximum velocity gradient. Absorption
is the case where the entire extinction is treated as
true absorption. We see that there is almost no
difference in maximum velocity in all these versions.
This is because the contribution of lines to the total
extinction in a hot medium is small. A great difference
is obtained only for ZR opacity, which underestimates
the true absorption on the photoeffect. According
to the Kirchhoff law, weaker absorption also means
weaker emission, a case closer to the adiabatic selfsimilar
one, whence the high ejecta velocity.

\begin{table}
\caption{Maximum matter velocity for various versions of
the series 14E at an ejecta mass of 14.7 \Msun
\label{velasym}}
\begin{center}
\begin{tabular}{rrr}
\hline
\hline
 \\[-3mm]
$E_{\rm kin}, \; 10^{51}$ erg & $R_0$, \Rsun & $u_{\max}, 10^3$km/s  \\
\hline
 \\[-3mm]
   1.6 &   48 &   42    \\
   1.3 &   48 &   37   \\
   1.0 &   48 &   33   \\
   0.7 &   48 &   28   \\
\hline
   1.2 &   10 &   59   \\
   1.3 &   48 &   37   \\
   1.1 &   300 &   20   \\
\hline
\end{tabular}
\end{center}
\end{table}

\begin{table}
\caption{Maximum matter velocity in various approximations
for opacity in the model 14E1.3 at an ejecta mass of
  14.7 \Msun
 \label{velopa}}
\begin{center}
\begin{tabular}{rrrr}
\hline
\hline
 \\[-3mm]
Opacity       &  $E_{\rm kin}, \; 10^{51}$ erg & $ R_0$, \Rsun
                 & $u_{\max}, 10^3$km/s \\
\hline
 \\[-3mm]
Standard    &   1.3 &   48 &   37.2   \\
`high'      &   1.3 &   48 &   37.7   \\
absorption  &   1.3 &   48 &   37.7   \\
 ZR         &   1.3 &   48 &   60     \\
\hline
\end{tabular}
\end{center}
\end{table}

 \subsection{The influence of a frequency grid
and approximations in treating the Compton effect}

It is particularly interesting to trace whether high
temperatures can be reached when more realistic
opacities than those in the ZR case are taken into
account for a more careful allowance for the photon
production, absorption, and scattering processes. In
particular, the cooling of electrons through the inverse
Compton effect (see, e.g.
\citet{Zeldovich1975}) should be
taken into account:
\begin{equation}
 W_{\rm CS} =\frac{4\sigma_T n_e k_{\rm B} T_e }{m_e c} U_{\rm rad} .
\label{compS}
\end{equation}

The results presented here for the models 14E took
this effect into account roughly --- in the approximation
of complete photon thermalization even at the
first scattering, i.e., the photons were assumed to
be absorbed with a cross section
 $\sigma_T k_{\rm B} T_e / {m_e c^2}$.
To
check whether this treatment leads to an underestimation
of the temperature, we carried out the following
experiment. Among the models for SN~1987A, we performed a series of calculations for the 14E1X2-
type versions, where such thermalization of photons
was switched off. Here, we took into account only
the coherent Thomson scattering, while the photon
energy could change only through the divergence of
bulk motion, just as in \citet{BlandfordPayne1981b} through the term:
\begin{equation}
  \frac{1}{3}  (\nabla \cdot \mathbf{u}) \nu \frac{\partial f}{\partial \nu},
\label{bulk}
\end{equation}
where $\nu$ is the photon frequency, $\mathbf{u}$ is the matter
velocity, and
 $f$ is the mean photon occupation number
(to be more precise, the zero angular moment of the
occupation number). In all of the remaining versions,
this term, of course, was also present in the equation
for the photon energy density.

%%%%% Analytic comparison of STELLA/Mihalas and Blandford & Payne
%\newpage

Let us compare in details the equations of \citet{BlandfordPayne1981a} with
equations used in our calculations.

The monochromatic radiation energy equation and
monochromatic radiation momentum equation with the accuracy $O(v/c)$ is written as
\citep[eq. 95.18-95.19]{Mihalas1984}:
\begin{eqnarray}
&&    \frac{DE_0(\nu_0)}{Dt}+\frac{1}{r^2}\frac{\partial}{\partial r}[r^2F_0(\nu)]+
    \frac{v}{r}[3E_0(\nu_0)-P_0(\nu_0)]+ \nonumber \\
&&    +\frac{\partial v}{\partial r}[E_0(\nu_0)+P_0(\nu_0)]+\frac{2a}{c^2}F_0(\nu_0)- \nonumber \\
&&    -\frac{\partial}{\partial\nu_0}\Big[\nu_0 \Big(
      \frac{v}{r}[E_0(\nu_0)-P_0(\nu_0)]+\frac{\partial v}{\partial r}P_0(\nu_0)+ \nonumber \\
&&      +\frac{a}{c^2}F_0(\nu_0)   \Big) \Big]
    =4\pi\eta_0(\nu_0)-c\chi_0(\nu_0)E_0(\nu_0)
\label{mdensity}
\end{eqnarray}
\begin{eqnarray}
&&    \frac{1}{c^2}\frac{DF_0(\nu_0)}{Dt}+\frac{\partial P_0(\nu_0)}{\partial r}
    +\frac{3E_0(\nu_0)-P_0(\nu_0)}{r}+ \nonumber \\
&&    +\frac{2}{c^2}\Big(\frac{\partial v}{\partial r} + \frac{v}{r}\Big)F_0(\nu_0)+
      \frac{2a}{c^2}[E_0(\nu_0)+P_0(\nu_0)]- \nonumber \\
&&    -\frac{\partial}{\partial\nu_0}\Big[\nu_0 \Big(
      \frac{v}{c^2r}[F_0(\nu_0)-Q_0(\nu_0)]+\frac{1}{c^2}\frac{\partial v}{\partial r}Q_0(\nu_0)+ \nonumber \\
&&      +\frac{a}{c^2}P_0(\nu_0)   \Big) \Big]
    =-\frac{1}{c}\chi_0(\nu_0)F_0(\nu_0)
\label{mmomentum}
\end{eqnarray}
Here $E_0(\nu_0),F_0(\nu_0),P_0(\nu_0),P_0(\nu_0)$ -- moments of the radiation field,
the affix ``0'' denotes the comoving frame,
$v$ -- the fluid velocity, $\chi$ -- the extinction coefficient,
$\eta$ -- the emission coefficient,
the time derivatives $(D/Dt)$ are evaluated in a moving fluid element.

{\sc stella} solves the equations, which are equivalent to equations
(\ref{mdensity}-\ref{mmomentum}), with the neglect of several terms \citep{Mihalas1984,Castor1972}:
\begin{eqnarray}
&& {\partial {{\cal J_\nu}}\over \partial t} =
   -{c\over r^2}\cdot {\partial \over
   \partial r}(r^2{\cal H_\nu})+c(\bar\eta_\nu-\chi_{\rm a} {\cal
   J_\nu})+ \nonumber \\
&& +{u\over r}(3{\cal K_\nu}-{\cal J_\nu})
  -  {1\over r^2}\cdot
   {\partial\over \partial r}(r^2u)
   ({\cal J_\nu}+{\cal K_\nu}) \nonumber \\
&& -{1\over \nu^3}
   \cdot {\partial\over \partial \nu} \nu^4\biggl[{u\over r} (3{\cal K_\nu}-{\cal
   J_\nu})
    -{1\over r^2} \cdot {\partial\over \partial r} (r^2u){\cal K_\nu}
   \biggr]
\label{comov}
\end{eqnarray}

\begin{eqnarray}
&& {\partial {{\cal H_\nu}}\over \partial t} =
-c{\partial {\cal K_\nu}\over \partial r}-
{c\over r}(3{\cal K_\nu}-{\cal J_\nu})-\nonumber\\
&& -2\biggl({u\over r}+{\partial u\over \partial r}
\biggr){\cal H_\nu}-c(\chi_{\rm a}+\chi_{\rm s}) {\cal H_\nu}
+\dot{\cal H}_{\nu_{\rm diff}} \;,
\label{impul}
\end{eqnarray}
where $u$ -- the fluid velocity, $\chi_{\rm a}=\chi_{\rm a}(\rho,T,\nu )$
-- true absorption coefficient, $\chi_{\rm s}$ -- scattering coefficient,
${\cal J_\nu, H_\nu, K_\nu}$ -- moments of the photon occupation number
and, for example, $E_\nu=8\pi h\nu^3/c^3{\cal J}_\nu$, ${\cal H}_{\nu_{\rm diff}}$ --
numerical stability term.

% Diffusion limit
In the diffusion limit when ${\cal{K}}\rightarrow {\cal{J}} /3$ at photon mean free path
$\lambda_p \rightarrow 0$
the equations (\ref{comov}-\ref{impul}) lead to:
\begin{eqnarray}
&& \frac{\partial {\cal J}_\nu}{\partial t} - \frac{\nu}{3}{1\over r^2}\cdot
   {\partial\over \partial r}(r^2u) \frac{\partial {\cal J}_\nu}{\partial\nu} =  \nonumber \\
&& = {1\over r^2}\cdot {\partial \over
   \partial r}\Big(r^2{\frac{c}{3(\chi_{\rm a}+\chi_{\rm s})}\frac{\partial {\cal J}_\nu}{\partial r}}\Big)
  +c(\bar\eta_\nu-\chi_{\rm a} {\cal
   J_\nu}),
\label{stellaEq}
\end{eqnarray}
which is equivalent to equation derived by \citet[eq. 97.68]{Mihalas1984} for the diffusion limit:
    \begin{eqnarray}
&& \rho\Big[\frac{D}{D t}\Big[\frac{ E_0(\nu_0)}{\rho}\Big]-
    \frac{\nu_0}{3}
     \Big(E_0(\nu_0)
     -\frac{\partial}{\partial \nu_0}[\nu_0 E_0(\nu_0)] \Big)  \nonumber \\
&&    \frac{\partial E_0(\nu_0)}{\partial \nu_0}%[\nu_0 E_0(\nu_0)] \Big)
     \frac{D}{D t}\Big(\frac{1}{\rho}\Big)
    \Big]%= \nonumber \\
 =\nabla\cdot\Big[\frac{c}{3\chi_0(\nu_0)}\nabla E_0(\nu_0) \Big] +
\nonumber \\
&& + \kappa_0(\nu_0)[4\pi B(\nu_0,T)-cE_0(\nu_0)]
\label{mihalasEq}
\end{eqnarray}
The left and right part of the equation (\ref{stellaEq}) correspond to the left and right
part of the equation (\ref{mihalasEq}) and it is supposed that
$\chi_{\nu}=\kappa_{\nu}+\sigma_{\nu}$ and
$\eta_{\nu}=k_{\nu}B_{\nu}+\sigma_{\nu}J_{\nu}$.
 Here $J_{\nu}$ is mean intensity or zeroth moment of radiation field over angles and $E_{\nu}=(4\pi/c)J_{\nu}$,
$B_{\nu}$ - isotropic specific intensity in thermal equilibrium,
 $\eta$ - emission coefficient, $\chi$ - extinction coefficient,
$k_{\nu}$ - true absorption coefficient, $\sigma$ - scattering coefficient,
$\rho$ - density of the matter, and recalling the equation of continuity
$D/Dt=\partial/\partial t+v\partial/\partial r$
\begin{equation}
   (D\ln\rho/Dt)=-r^{-2}[\partial(r^2v)/\partial r]=-(\partial v)/(\partial r)-(2v/r).
\end{equation}

The equation used by \citet[eq.18]{BlandfordPayne1981a} can be rewritten as follows
(we omit the photon energy redistribution by Compton scattering due to
{\sc stella} does not take it into account):
\begin{eqnarray}
 \frac{\partial \bar{n}}{\partial t}+u\cdot\nabla \bar{n} -
    \frac{1}{3}(\nabla\cdot u)\nu\frac{\partial \bar{n}}{\partial \nu} =
    \nabla\cdot\Big(\frac{1}{3n_e\sigma(\nu)}\nabla \bar{n}\Big)
%         + \nonumber \\
%    &&    +\frac{1}{\nu^2}\frac{\partial}{\partial\nu}
%        \Big[\frac{n_e\sigma(\nu)}{m_e}\nu^4\Big(\bar{n}+
%        T\frac{\partial \bar{n}}{\partial\nu}\Big)\Big]
    +\bar{j},
\label{blandfordpayne}
\end{eqnarray}
where $\bar{n}$ - mean occupation number, i.e. $\cal J$ in the equation
(\ref{stellaEq})

The first two terms of the equation (\ref{blandfordpayne}) correspond to the first term
of equation (\ref{stellaEq}) (the time derivative in {\sc stella} is taken at fixed Lagrangean radius).
The third term of (\ref{blandfordpayne}) corresponds to the second term of equation (\ref{stellaEq})
and describes the heating of the radiation through compression.
The first term of the equation (\ref{blandfordpayne}) in the right part
corresponds to that one in the equation (\ref{stellaEq}) and describes diffusion of photons in space.
The photon source term $\bar{j}$ in equation (\ref{blandfordpayne}) is added without without any
corresponding absorption term and it is inconsistent with thermodynamics \citep{PsaltisLamb1997}.
The equation (\ref{stellaEq}) contains both emission and absorption terms:
$c(\bar\eta_\nu-\chi_{\rm a} {\cal J_\nu})$.

We note that our equations include all terms of final equation of \citet{BlandfordPayne1981a},
but our approach is more accurate because all variables are considered in Lagrangean frame,
while
\citet{BlandfordPayne1981a} did not distinguish clearly the fluid and the inertial frame
\citep{Fukue1985}.
Moreover, we do not add any inconsistent terms and do not confine ourselves to the diffusion limit.

%%%%%%%%%%%%%%%%%
The characteristic time scales of the changes in spectrum due
to Compton scattering $t_h^{-1},t_c^{-1}$,
and the divergence of bulk motion
$t_u^{-1}$ are, respectively,
\begin{equation}
t_h^{-1} \sim n_e \sigma_T c \Big( \frac{4kT_e}{m_e c^2} \Big),
\end{equation}
\begin{equation}
t_c^{-1} \sim n_e \sigma_T c \Big( \frac{h\nu}{m_e c^2} \Big),
\end{equation}
\begin{equation}
t_u^{-1} \sim (\nabla \cdot \mathbf{u})  \sim n_e \sigma_T c \Big(\frac{1}{3} \frac{u^2}{c^2} \Big),
\end{equation}
where the scale length $\sim c/(n_e \sigma_T u)$ \citep{BlandfordPayne1981a}.
In the case of ultrarelativistic motions
(see the transfer equation (\ref{mihtraneq})
below in the comoving
frame of reference), the term with $\nu {\partial f}/{\partial \nu}$ takes a
more complex form (the intensity is $I=2h \nu^3f/c^2$).

Another difference of the 14E1X2-type versions
is a wider frequency grid: the number of frequency
bins was doubled and the minimum wavelength $\lambda$  was
set equal to $\lambda$ was
set equal to $10^{-2}$ \AA\ (instead of $1$ \AA\ in the standard series 14E).

The resulting outburst light curves for two versions
of our calculation are presented in Fig.~\ref{14e1lumb}.
We see that the difference is very small.
Fig.~\ref{cadr14e1} shows typical spectra at the epoch of shock breakout
for 14E1X2. Increasing the grid shows that the shock
breakout in X rays becomes visible much earlier than
in visible light, because the absorption is weaker
there. However, the fluxes in the hard part of the
spectrum remain low. The figure also shows the best
fit by a blackbody spectrum with a temperature that
we call the color one.

\begin{figure}
\centering
\includegraphics[width=84mm] {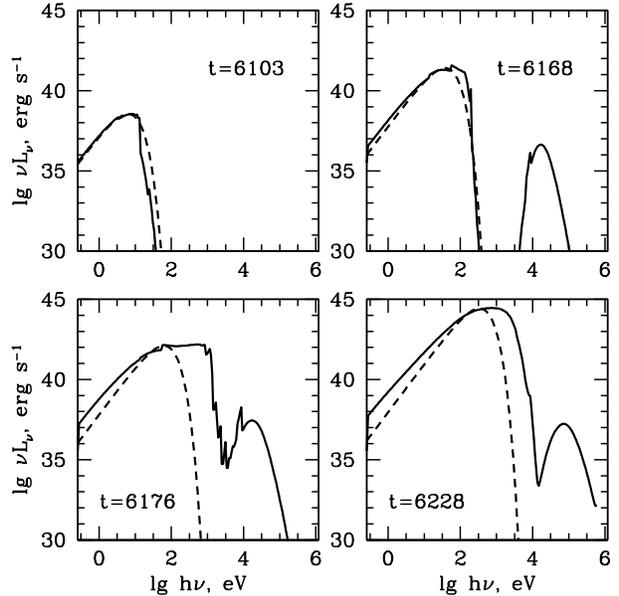}
\caption{Spectral flux distributions $\nu L_\nu$ for four instants of time (shown in seconds)
          for the variant 14E1X2 at shock breakout
          (solid line) and the best fit by a blackbody spectrum(dashed line).
          The time lag was disregarded.
}
\label{cadr14e1}
\end{figure}

 \subsection{The hardest semirelativistic variant, the SN Ib/c model}

As we see from Table \ref{velasym}, the maximum ejecta velocity
depends strongly on presupernova radius: it is
higher in more compact stars. Therefore, it is interesting
to consider the shock breakout for type Ib/c
supernovae (their progenitors are Wolf--Rayet stars
with radii of the order of the solar one or even smaller).
The type-Ib/c presupernova model that we use was
constructed with the {\sc kepler} code by an evolutionary
computation from a main-sequence star in
\citet*{Woosley1995}, the model $7A$.
At the end of its
evolution, the presupernova star has a core composed
of helium and heavy elements with amass of
$3.199M_{\odot}$ and a radius of $1.41\cdot 10^{11}$ cm.
The radius in this case
was fixed ``manually'', because in the outer stellar layers
{\sc kepler} models the stellar wind and the model is
not in hydrostatic equilibrium.

%ALTO This needs to be corrected
This model
was chosen because the velocity of matter at shock breakout
reaches the values about a half of speed of light. At such velocities, the relativistic
corrections to the radiative transfer are significant and
applying the {\sc rada} code here is of great interest.
Although {\sc stella} does not work with hydrodynamic
flows with large Lorentz factors, it takes into
account the relativistic effects with an accuracy of
$(v/c)$ sufficient for the models we consider.
The
results of calculations of light curves and spectra of this model, which we denote s1b7a2,
in the variant of the algorithm {\sc rada} with a soft spectrum ($\lambda>1$\AA) are the following.

Figure \ref{lcIbc}
% and \ref{spIbc}
compare the calculated light curves for the model.

\begin{figure}
%\centering
{\includegraphics[width=84mm]{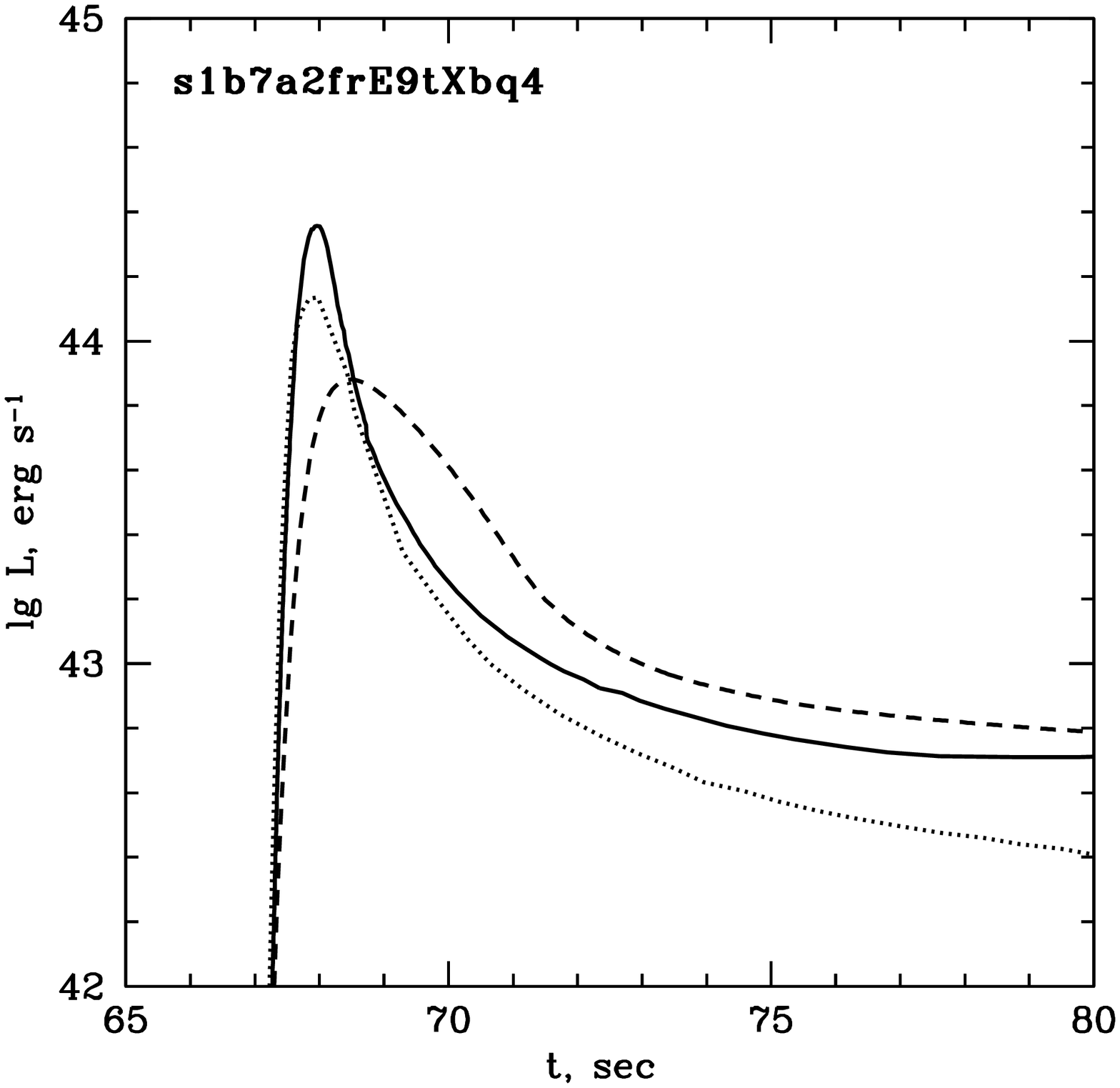}}
{\includegraphics[width=84mm]{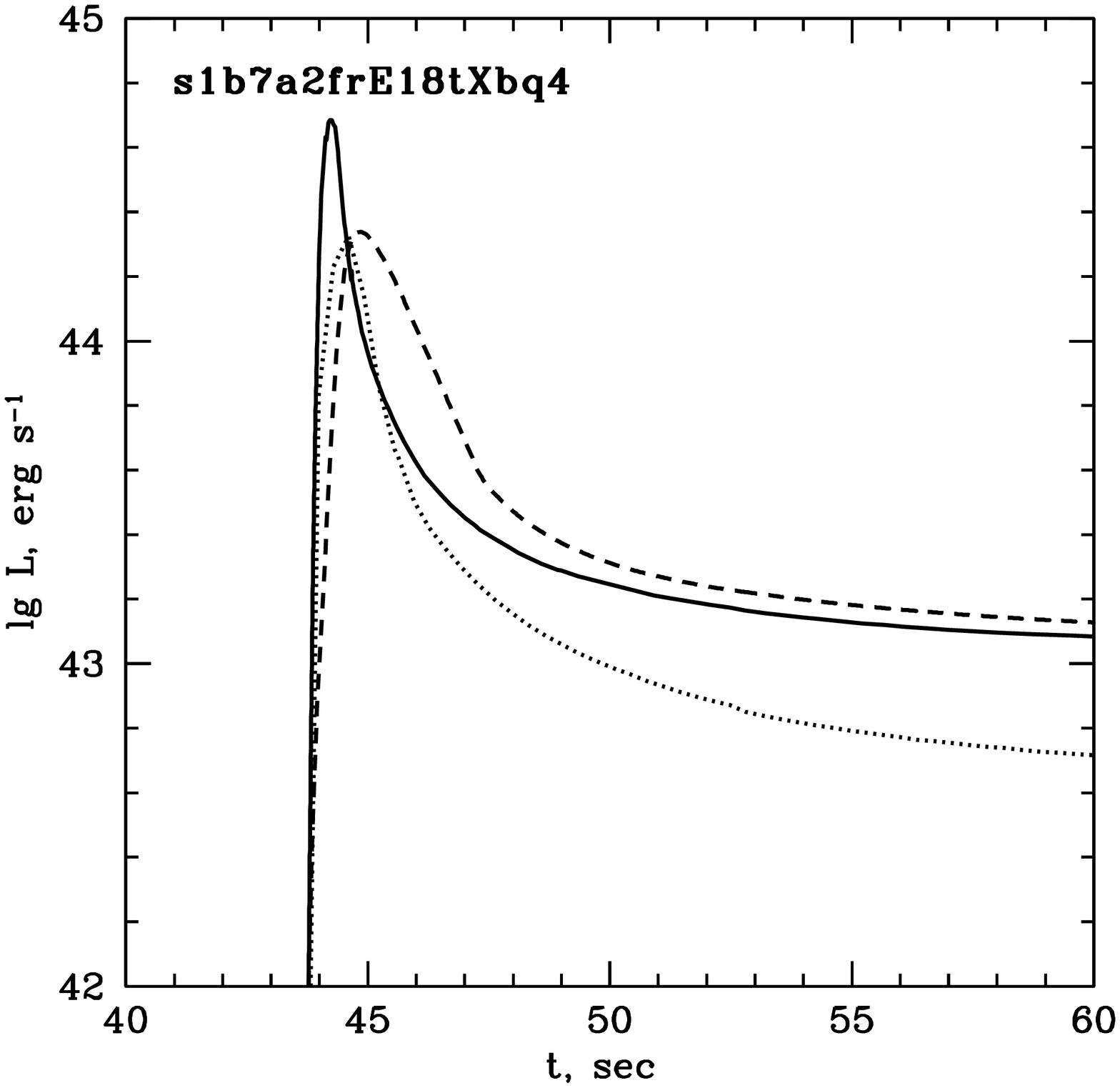}}
\caption{{\color{black}Comparison of the bolometric light curves at the epoch of shock breakout
         for a type Ib/c presupernova models: explosion energy $E=9\cdot10^{50}$~erg, 
         maximum matter velocity $v_{max}\approx 0.3 c$ (top) 
         and $E=1.8\cdot10^{51}$~erg, $v_{max}\approx 0.5 c$ (bottom).
         The solid and dotted lines represent,
         respectively, the {\sc stella} and {\sc rada} calculations
         in comoving frame of reference.
         %which do not take into account radiation time delay.
         Dashed line represents {\sc rada} calculations in observer's frame of reference
         taking into account radiation time delay
         in the observer's frame of reference.}
}
\label{lcIbc}
\end{figure}

%\begin{figure}
%\centering
%\includegraphics[width=84mm]{lcIbc2.ps}
%\caption{Comparison of the monochromatic light curves at the epoch of shock breakout
%         in the observer's frame of reference for
%         a type Ib/c presupernova model. Between the 81st and 82nd seconds,
%         a break (steep decline) of the light curve attributable to
%         the relativistic delay effect is observed. The thin and thick lines represent,
%         respectively, the {\sc stella} calculation with the delay
%         effect and the {\sc rada} calculation.
%         The number N corresponds to the pair of light curves below it.
%}
%\label{lcIbc}
%\end{figure}

%\begin{figure}
%\centering
%\includegraphics[width=84mm]{spIbc2.ps}
%\caption{Comparison of the instantaneous spectra for various times t at
%         the epoch of shock breakout in the observer's frame of
%         reference for a type Ib/c presupernova model. The thin and thick lines
%         represent, respectively, the {\sc stella} calculation with
%         the delay effect and the {\sc rada} calculation.
%         The number N corresponds to the pair of spectra closest to it.
%}
%\label{spIbc}
%\end{figure}

We see from the computational data that allowance
for the delay effect and a strict allowance
for the relativistic radiative transfer affect the light
curve shape. It leads to decrease in radiation flux at maximum and
to a broadening of the light curve peak.
The effect of geometrically eclipse the radiation of the outburst
from the edge of the star (see Fig.~\ref{geom5}d in Appendix B)
is revealed in the calculations but it is quite small in this model
due to small velocities after shock breakout (0.2-0.5 c) and
can not be seen on the graph in details.
The light curve presented here is calculated more accurately than
we discussed earlier \citep{Tolstov2010}.

Below we consider s1b7a2
variants of {\sc stella} with a hard spectrum ($\lambda<1$\AA) as well.
Options differ from that described above, only by the numerical
treatment of radiation transfer.

{\color{black}
Program {\sc stella}, which is based on the nonrelativistic equations of hydrodynamics,
can be modified to allow for relativistic velocities. Instead of the velocity $u$, the numerical scheme may work
with the quantity $U \equiv \gamma u$.
The equations for $U$ appear
almost as in the Newtonian limit for $u$ (see, e.g., \citet{Urzhumov2002})
but the velocity $u=U/(U^2/c^2 +1)^{1/2}$ now can not exceed the speed of light. In the current work we do not use this
modification and limit ourselves to the case when $u < c/2$.
}

In all the versions of the model s1b7a2, described
above, we changed neither the explosion energy nor
the radius. The variant s1b7a2X was computed
by {\sc stella} on the same spatial grid as s1b7a2, but the number of frequency bins was doubled and
the minimum wavelength $\lambda$ was set equal to $10^{-3}$ \AA\ (instead of $1$ \AA\ in s1b7a2).
In this case, no photon
thermalization under the Compton effect was imposed.

In Fig.~\ref{tpms1b7}, the matter temperature for the variant s1b7a2X
is plotted against the Lagrangean
mass $M_r$ measured from the surface.
We see that
the maximum temperatures are enormous --- up to $10^{10}$~K (i.e., of the order of MeV).
The spectrum for
an observer in the comoving frame at the surface
is shown in Fig.~\ref{cadrs1b7}.
Since the {\sc stella} algorithm includes the evolution
of photons in a converging flow in the shock in
the same approximation (\ref{bulk}), as
considered by \citet{BlandfordPayne1981b}, one could think that
our computation confirms their analytical result.

\begin{figure}
\begin{center}
\includegraphics[width=84mm] {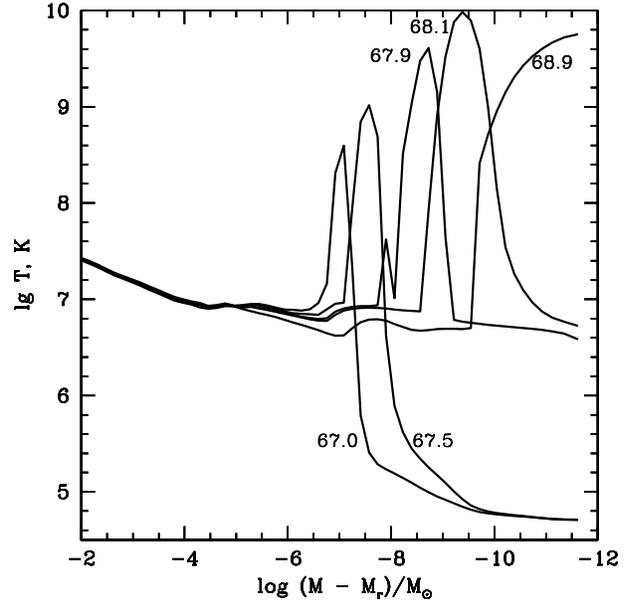}
\caption{Matter temperature for the variant s1b7a2X
         at shock breakout versus Lagrangean mass $M_r$ measured from the surface.
         The time in seconds is given near the curves. The temperature peak is at an optical depth
         $\tau \sim 200, \; 50, \; 4, \; 1, \; 0$ at times 67.0, 67.5, 67.9, 68.1, 68.9 s.
\label{tpms1b7}
}
\end{center}
\end{figure}

\begin{figure}
\begin{center}
\includegraphics[width=84mm] {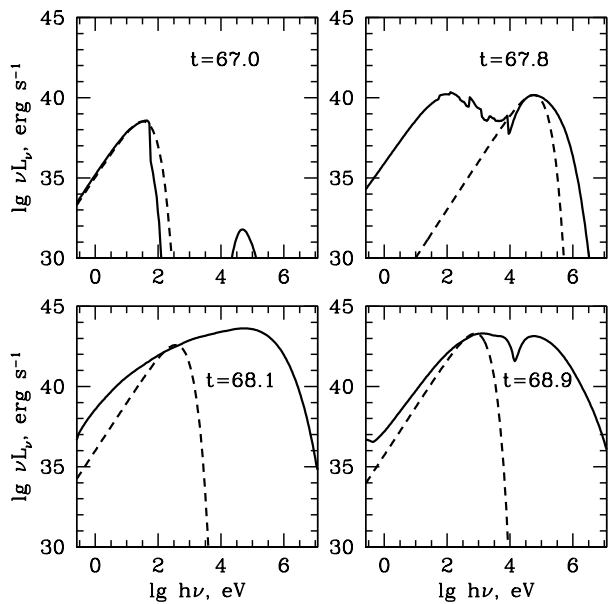}
%\plotone{tpmbq2.ps}
\caption{Spectral flux distributions $\nu L_\nu$ for four instants of time
         (shown in seconds) for the variant s1b7a2X at shock breakout
         (solid line) and the best fit by a blackbody spectrum(dashed line).
         The time lag was disregarded.
\label{cadrs1b7}
}
\end{center}
\end{figure}

In fact, however, the high density and hardness of
the radiation in such a shock obtained in the s1b7a2X
computation require a more careful allowance for the
photon production, absorption, and scattering processes.
In particular, the cooling of electrons through
the Compton effect, as was discussed above for the
variant 14E1X2 as an example, should be taken into
account.

Nevertheless, it turns out that allowance even for
the weaker double Compton effect
\citep{MandlSkyrme1952,Weaver1976,Lightman1981} leads
to a sharp change of the results. We will postpone its
rigorous allowance until a future paper, while here this
effect was simulated by a very small admixture of true ``gray'' (i.e., frequency-independent)
absorption, $10^{-6}$ of the Thomson scattering in an SN~Ib progenitor.
We called this variant s1b7a2Xm6.

The plots in Figure \ref{tpms1b7m6} and \ref{cadrs1b7m6} show that the matter
temperature and spectrum hardness for the variant
s1b7a2Xm6 decrease sharply compared
to s1b7a2X.

\begin{figure}
\begin{center}
\includegraphics[width=84mm] {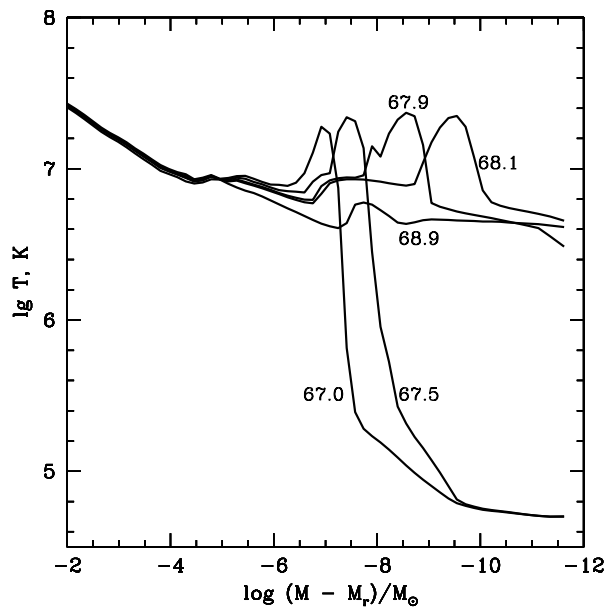}
\caption{Matter temperature for the variant s1b7a2Xm6
         at shock breakout versus Lagrangean mass $M_r$ measured from the
          surface. The time in seconds is given near the curves.
         The temperature peak is reached at $ \tau \sim 200 \, 50, \ 4, \; 0.5 $
         at times 67.0,
         67.5, 67.9, 68.1 s, respectively, and virtually disappears subsequently.
\label{tpms1b7m6}
}
\end{center}
\end{figure}

\begin{figure}
\begin{center}
\includegraphics[width=84mm] {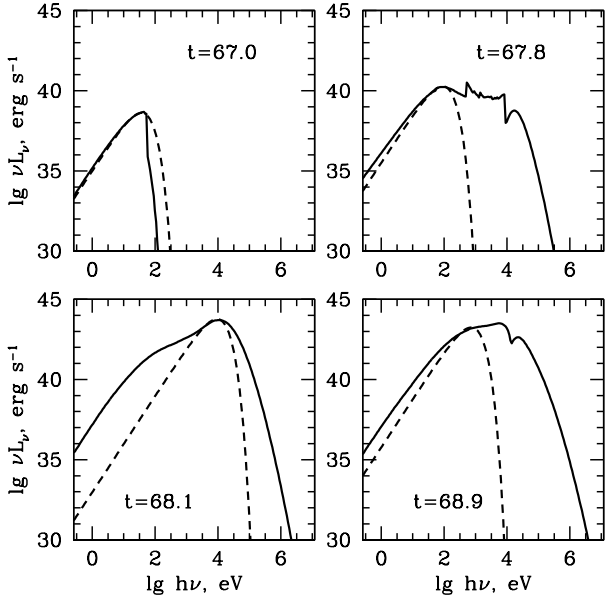}
\caption{Spectral flux distributions $\nu L_\nu$ for four instants of time
    (shown in seconds) for the variant s1b7a2Xm6 at shock
    breakout (solid line) and the best fit by a blackbody spectrum (dashed line).
    The time lag was disregarded.
\label{cadrs1b7m6}
}
\end{center}
\end{figure}

The previous results were presented for an observer
at the ``edge'' of the supernova ejecta in the
comoving frame. The {\sc stella} results can be carefully
transformed to the rest frame using the {\sc rada} algorithm.
%The results of this comparison for the models of SN ~ 1987A are Tolstov ~ (2010).
Here, Fig.~\ref{graph} compares the monochromatic light
curves at shock breakout in the reference frame of
an observer at rest for a type Ib/c presupernova.
The gray line represents the {\sc stella} computation
without the time lag. The black line represents the
computation in the observer's frame of reference with
a full allowance for the time lag (in the blackbody approximation
for brightness) computed by the {\sc rada} algorithm.

\begin{figure}
\begin{center}
\includegraphics[width=84mm]{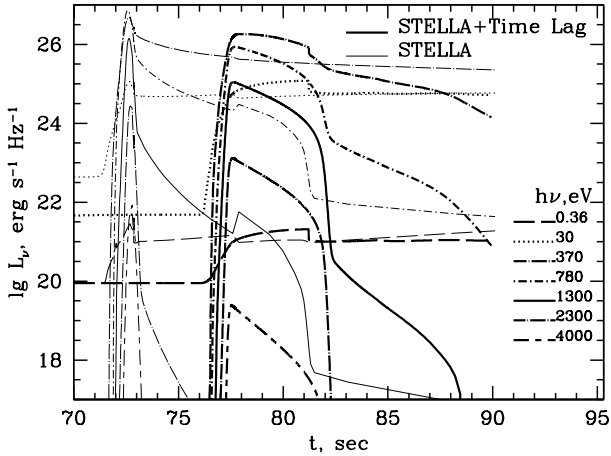}
\caption{Comparison of the monochromatic light curves at shock breakout for a type Ib/c
         presupernova.
         The gray line represents the {\sc stella} computation in the comoving frame of
         reference with the only correction for the observer's time $t\rightarrow t-R(t)/c$.
         The black line represents the computation in the observer's frame of reference with a full allowance for the
         time lag (in the blackbody approximation for brightness).
         %The number N corresponds to the pair of spectra closest to it. %todo
\label{graph}
}
\end{center}
\end{figure}

Note that the flux in the computation including
the time lag was calculated under the assumption
of blackbody intensity. This assumption allows the
aberration and Doppler effects to be properly taken
into account, but it leads to a difference in fluxes
before the shock breakout. On the other hand, the
calculation of fluxes by disregarding the time lag is
based only on the work with the radiation times.
Therefore, the comparison in the figure may be considered
only as a qualitative one, demonstrating the
main differences and peculiarities of the light-curve
shapes.

\section{Conclusions}

In this paper, we found that the high-temperature
peak behind the shock front and the possibility of the
formation of a hard power-law ``tail'' in the radiation
spectrum are suppressed through a very weak process
of photon absorption and production with a cross section
at a level of one millionth of the Thomson scattering
cross section. At this level, the absorption can
be provided by double Compton effect.

In theoretical models of type Ib/c supernovae shock breakout the motion of matter
at velocities near the speed of light should be taken into account.
Development of the numerical algorithm {\sc rada} for solving
radiative transfer with the hydrodynamic part of the algorithm {\sc stella} provides
a reasonable allowance for the relativistic effects.

Our calculations of supernova shock breakouts
can be used for evaluating and interpreting the detection of supernova
explosions in planned space experiments.
In the near future the number of observed outbursts will grow due to the launches
of new spacecrafts and the theoretical models are
called for because they can be used to predict the number and nature of
observed outbursts.

One of these experiments could be LOBSTER \citep{CalzavaraMatzner2004}, which
is seems to be discarded now.
Anyway, the characteristics of a detector of this type are
more favorable for the detection of outbursts from
presupernovae that are red and blue giants. For
Wolf-Rayet stars, the detection rate is estimated
to be several outbursts per year from a maximum
distance of $50$ Mpc. This distance is a natural limit
of the detection of outbursts of this kind,
because the daily detection limit of the spacecraft detector
would be about $10^{-12}$ erg cm$^{-2} s^{-1}$ \citep*{Priedhorsky1996}.
For events in our Galaxy, the outburst spectrum will be
quite resolvable and our computation of
model $7A$, already refined by relativistic effects, can be used for
comparison with the observational data.

One of the most interesting problems of the stellar
evolution theory is an unknown number of outbursts from
collapsing SN 1987A-type supernovae. They are difficult
to observe due to the short outburst duration
($\sim 100$ s), but the outburst brightness makes it possible
to detect them with X-ray detectors.

Type Ib/c supernova explosions are also difficult to
observe due to their short duration ($\sim 10$ s), but, in this
case, the radiation has the hardest spectrum and the
X-ray radiation from the shock propagating within
the presupernova wind can be detected. The detection
of type Ib/c explosions is very important for the theory
of hypernovae, since it will allow us to describe better
the connection of supernovae with gamma-ray bursts
and, in the case of nearby supernovae, there can be
a correlation between the events and the detection of
gravitational waves and neutrinos.

According to the
estimations of \citet{CalzavaraMatzner2004}, LOBSTER would have
the following possibility of outbreak registration for 3 years:
from $50$ to $600$ supernovae type II and some type Ib/c
within $250$ Mpc.
These estimations are based only on
several existing calculations taking into account the multigroup radiation transport
described in the literature -- the model SN 1993J
\citep{Blinnikov1998} and SN 1987A \citep{Blinnikov2000},
and the rest of the spectrum of data
obtained by scaling these calculations. To improve them, we can use
the theoretical modeling methods we developed.

A further development of numerical one-dimensional
supernova explosion models requires both a more
accurate allowance for the hydrodynamic effects
and a more complete description of Compton scattering of radiation.
The observed asymmetry effects require
developing multidimensional algorithms and radiative
transfer calculations, but the success  achieved  in
modeling supernova explosions serves as an impetus
for future studies.

\section{Acknowledgments}

We are grateful to V.~S. Imshennik, V.~P. Utrobin for useful discussions,
P.~V. Baklanov, E.~I. Sorokina for assistance in the development of the algorithm {\sc stella},
S. Woosley and K. Nomoto for providing models of presupernovae,
{\color{black}S.A.E.G. Falle for the valuable review comments}.
Part of the work was done at MPA (Garching, Germany) and IPMU (Kashiwa, Japan),
we thank W. Hillebrandt, R. Sunyaev,
H. Spruit, E. M\"uller and H. Murayama for their hospitality and support.

This work was partly supported by RFBR grants 10-02-00249-A, 10-02-01398-A,
Grants Research Schools 2977.2008.2, 3884.2008.2.,
Grant SNSF (Swiss National Science Foundation) program SCOPES No.~IZ73Z0-128180/1
and by Grant of the Government of the Russian Federation 11.G34.31.0047.

\bibliographystyle{mn2e}
\bibliography{bibfile}

\appendix
\section{Time-Delay-Spread Light Curves}

The light radiated by different places of stellar photosphere
at the same time $t$ reaches observer at different times spread
over time interval of width $R/c$, $R$ and $c$ being
the photosphere radius and speed of light, respectively.
The observer actually sees a convolved signal radiated by
photosphere within the time interval $\Delta t =R/c\,$.
Mathematically, the observed luminosity $L(t)$ can be expressed
through the intrinsic, not spread, luminosity $L_0(t)$
by the following integral:
\begin{equation}
   L(t) = 2\int_0^1\!\! L_0(t')\, x\, dx\, ,\qquad
   t'\equiv t-(R/c)\cdot (1-x)\, .\label{Lt}
\end{equation}
 This expression is based on 3 basic assumptions.
 First, it implies the distance $D$ between observer and the star
 is very large: $R/D\ll 1$. Second, it assumes that the
 radiation intensity $I$ is isotropic all over the photosphere ---
 i.e., $I$ does not depend on the angle $\theta$ between the direction
 of radiation propagation and the perpendicular to the photosphere surface.
 Third, the Eq.$\,$(\ref{Lt}) neglects the change in $R$ during
 the time interval $\Delta t =R/c$. This is a good approximation
 as long as the velocity of photosphere, $V=dR/dt$ (do not confuse
 with that of matter crossing the photosphere!) remains small in comparison
 with the speed of light, say $|V/c|\lsim 0.1$.
 This is true for the shock wave breakout in common supernovae
 of Types II, Ib, and Ic.

 \begin{table*}
 \begin{minipage}{126mm}
%\begin{center}
 \caption{The properties of the time-spread exponential light curve.}
 \label{Twidth}
%\vspace*{2mm}
\begin{tabular}{lllllllll}
%  after \\: \hline or \cline{col1-col2} \cline{col3-col4} ...
 \hline\hline
  & & & & & & & &\\[-3mm]
  $\alpha = R/(c\tau)$          & 0.0   & 0.2   & 0.4  & 0.6    & 1.0   & 2.0
                                                 & 4.0   & 8.0$\,^\ast$ \\
  $L_m/(E/\tau)$    & 1.0   & 0.884 & 0.794 & 0.722 & 0.614 & 0.451
                                                 & 0.299 & 0.181 \\
  $t_m/\tau$        & 0.0   & 0.182 & 0.336 & 0.470 & 0.693 & 1.10
                                                 & 1.61  & 2.20 \\
  $\Delta t_q/\tau$ & 0.921 & 1.07  & 1.22  & 1.36  & 1.63 & 2.29
                                                  & 3.67 & 6.33 \\
  $\lambda$         & 1.0   & 1.07  & 1.15  & 1.23  & 1.44 & 2.19
                                                  & 6.20 & 92.9 \\[1mm]
      \hline\hline\\[-2mm]
 \multicolumn{9}{l}{\footnotesize $^\ast\;$In the limit
  $\alpha\rightarrow\infty$ (see \fref{Fdelta}):
  $\;L_m\frat{cE}{R}=2$, $\; \frat{t_m c}{R}=0$,
  $\; \Delta t_q =0.6\,\tau\alpha= 0.6\frat{R}{c}$.}
\end{tabular}
%\end{center}
 \end{minipage}
\end{table*}

 Integrating both sides of Eq.$\,$(\ref{Lt}) by $t$, one can
 make certain that $\int^\infty_0L(t)dt\\
 =\int^\infty_0L_0(t)dt$.
 Thus as it should be, the transformation in question conserves the
 total radiated energy --- it only turns out to be redistributed in time for
 a remote observer. In a trivial limit $R\rightarrow 0$ (no time delay
 at all), Eq.$\,$(\ref{Lt}) gives, naturally,
 \mbox{$L(t)=L_0(t)$.}
 The following two simple but instructive examples
 illustrate how Eq.$\,$(\ref{Lt}) actually works.

 \subsection*{Useful Examples}
 Consider the case when the intrinsic luminosity time-scale
 $\tau$ is very short as compared to $R/c$,
 or $\alpha\equiv R/(c\tau)\rightarrow\infty$. In such a limit
 $L_0(t)$ can be represented as the $\delta$-function:
 $L_0(t)=E\,\delta(t)$, where $E$ is the total
 radiated energy. Inserting this $L_0$ in Eq.$\,$(\ref{Lt})
 we obtain
 \begin{equation}
 L(t)=\left\{\begin{array}{ll}
   0\, , & t\le 0\, , \\
   \frat{c}{R}\, E\left(1-\frat{ct}{R}\right),&
   0<t\le\frat{R}{c}\, ,\\
   0\, , & t>\frat{R}{c}\, .
 \end{array}\right.\label{Ldelta}
 \end{equation}
 The resulting convolved luminosity $L(t)$ has a triangle form
 as shown in \fref{Fdelta}.

\begin{figure}
\begin{center}
\includegraphics[width=50mm]{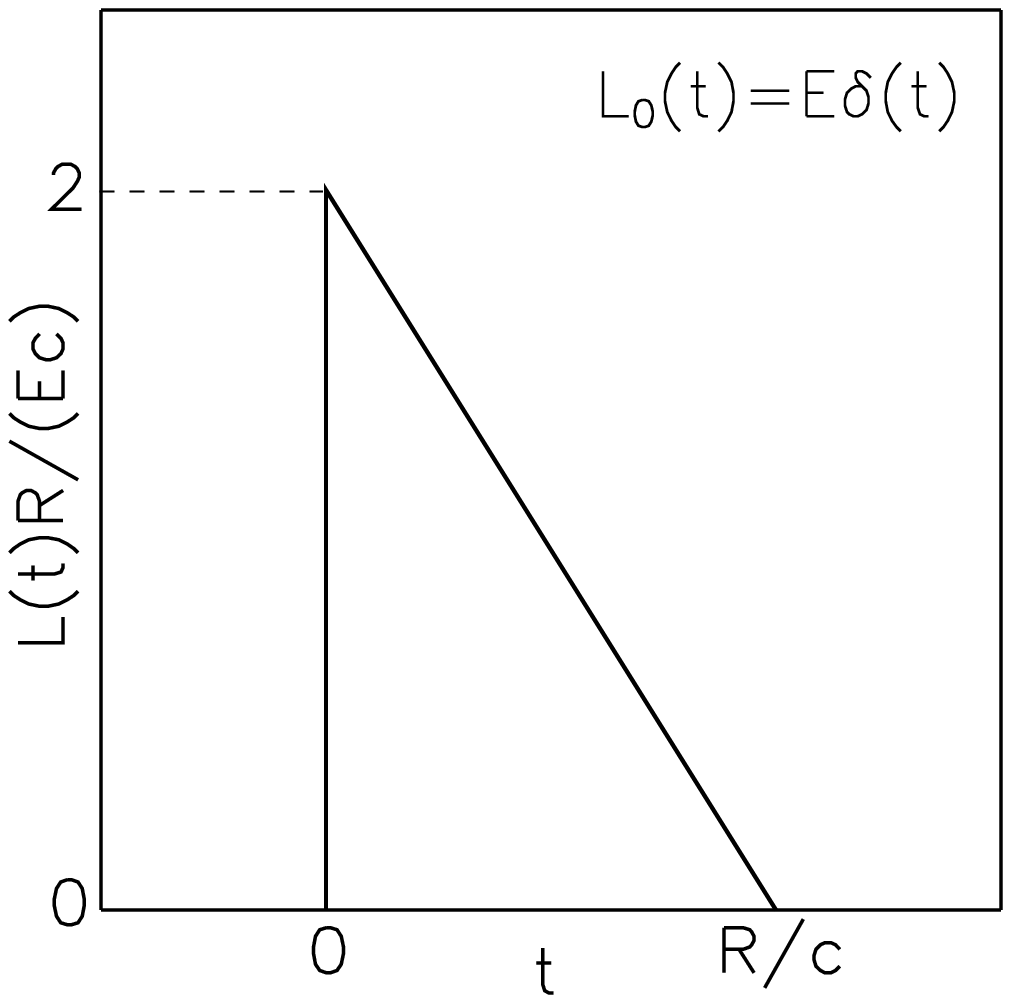}
\caption{Time-delay-spread $\protect\delta$-function.
\label{Fdelta}
}
\end{center}
\end{figure}

 Another example providing the integral in Eq.~(\ref{Lt}) in
 a closed form is the exponential function:
 \begin{equation}\label{Lexp0}
 L_0(t)=\left\{\begin{array}{ll}
   0\, , & t\le 0\, , \\
 \frat{E}{\tau}\,\exp (-t/\tau)\, , & 0< t < \infty\, .
 \end{array}\right.
\end{equation}
The result is:
 \begin{eqnarray}
 && \!\!\!\! \!\!\!\! L(t)=2\,\frat{E}{\tau}\!\left(\frat{c\tau}{R}\right)^2\times\\
 && \times
 \left\{\begin{array}{ll}
   0\, , &\!\!\!\!\!\!\!\!\! t\le 0\, , \\
  \Big[\!\left(1+\frat{R}{c\tau}\right)
  \!\!\left[1-\exp\!\left(-\frat{t}{\tau}\right)\right]
   -\frat{t}{\tau}\Big], &\!\!\!\!\!\!\!\!\! 0<t\le\frat{R}{c}\, ,\\
                          & \\[-3mm]
   \left[\!\left(1+\frat{R}{c\tau}\right)
   \!\!\Big[1-\exp\!\left(-\frat{R}{c\tau}\right)\right]
   -\frat{R}{c\tau}\Big]\!\times & \\
   \times
   \exp\left(\frat{R}{c\tau}-\frat{t}{\tau}\right)
   \, , &\!\!\!\!\!\!\!\!\! \frat{R}{c}<t<\infty\, .
 \end{array}\right.\nonumber\label{Lexpd}
 \end{eqnarray}
 Figure$\,$\ref{Fexpd} shows $L(t)$ for a number of values of
 parameter $\alpha = R/(c\tau)$.
%\begin{figure}
%\begin{center}
%\includegraphics[width=50mm]{Triangqz.eps}
%\caption{Time-delay-spread $\protect\delta$-function.
%\label{Fdelta}
%}
%\end{center}
%\end{figure}
\begin{figure}
\begin{center}
\includegraphics[width=70mm]{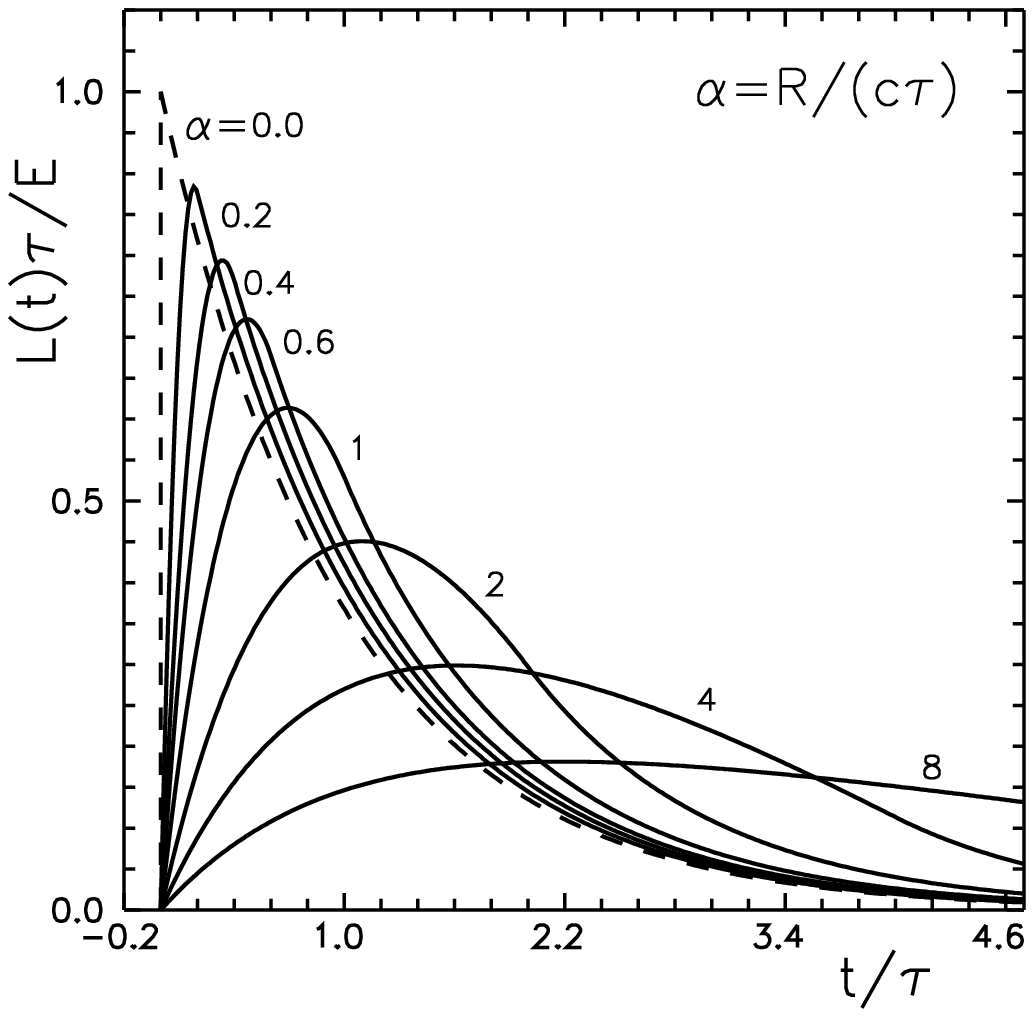}
\caption{Time-delay-spread exponential function.\protect\\
  Shown by a dashed curve is
  the initial $L_0(t)$ (Eq.$\protect\,$\protect\ref{Lexp0})
  that corresponds to $\protect\alpha=0$.
\label{Fexpd}
}
\end{center}
\end{figure}

The next \fref{Fexpdr} shows $L(t)$ in terms of $Ec/R$ as a function
of time in terms of $R/c$. The dashed curve represents
the $\delta$-function response (Fig.$\,$\ref{Fdelta}) in a limit
$\alpha\rightarrow\infty$. We see that the curve for $\alpha=8$
is already not so far from this limit.

\begin{figure}
\begin{center}
\includegraphics[width=70mm]{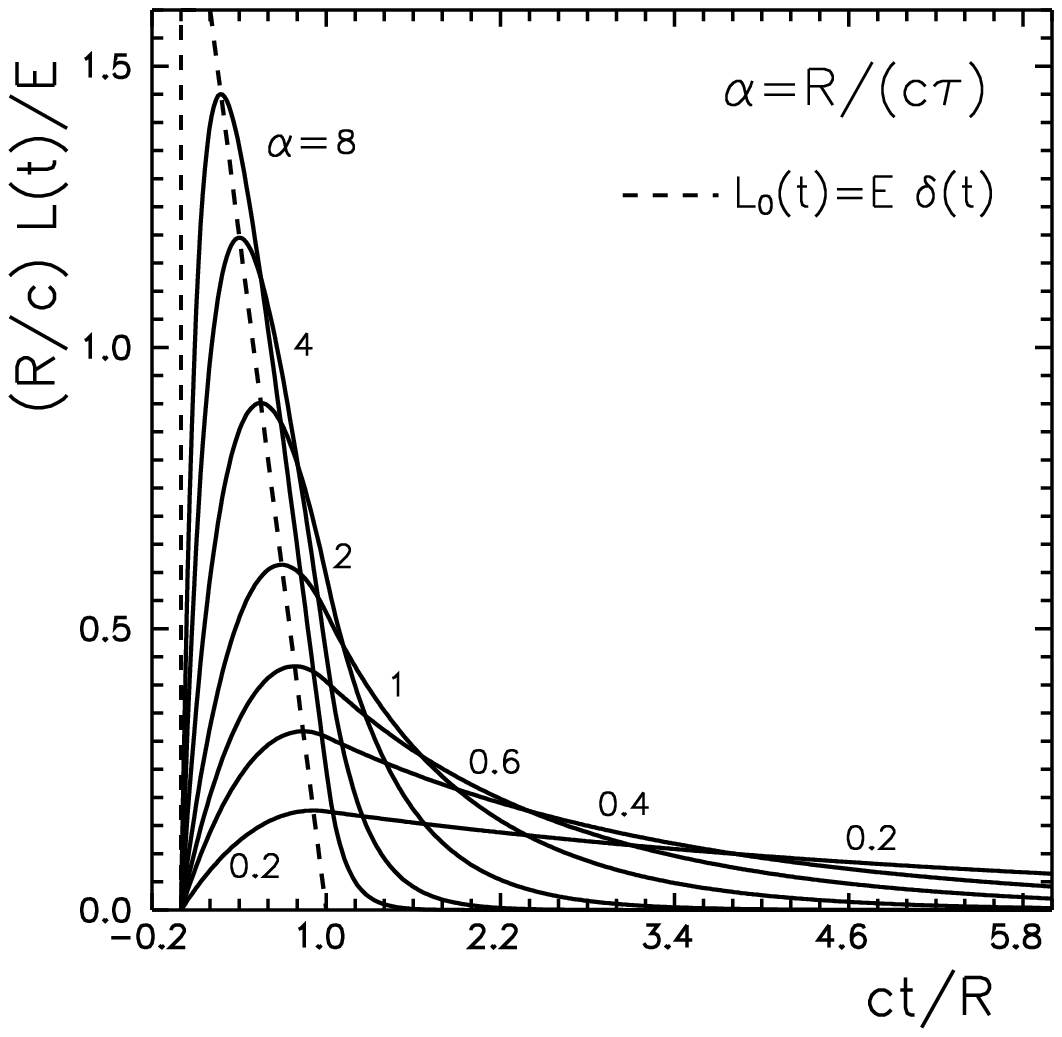}
\caption{Same as in \protect\fref{Fexpd} but for the luminosity
  and time in terms of $Ec/R$ and $R/c$, respectively.
\label{Fexpdr}
}
\end{center}
\end{figure}

 It is easy to find that
 the time-spread light curve (Eq.$\,$\ref{Lexpd}) attains a maximum
 $L_m$ at $t=t_m$ within the time interval $0< t < R/c\,$:\\
 \begin{equation}\label{Lmax}
 L_m\, =\,\frat{2}{\alpha^2}\,\frat{E}{\tau}\,
 \left[\alpha -\ln(1+\alpha)\right],\quad
 t_m\, =\,\tau\ln(1+\alpha)\, .
\end{equation}
 To find the light curve width $\Delta t_q$ at a given level
 $L(t)/L_m = q < 1$ one has to specify $\alpha$
 and solve a couple of transcendent equations.
 Table$\,$\ref{Twidth} presents $\Delta t_q$ for $q=10^{-0.4}$
 (one stellar magnitude below maximum) along with $L_m$ and $t_m$
 for the values of $\alpha$ shown in
 Figs.$\,$\ref{Fexpd} and \ref{Fexpdr}.

 The dependence of $\Delta t_q$ on $\alpha$ can be approximated
 with a good accuracy (better than $\approx 2$\%)
 by the following simple formula:
 \begin{equation}\label{Dtalph}
  \Delta t_q\, =\,\tau\left[0.921+0.6\,\alpha
   \left(1+\frat{0.45}{1+\sqrt{\alpha}}\right)\right] ,\;
   0\le\alpha<\infty\, .
\end{equation}

 Finally, it is worthy to notice that for time $t > R/c$
 the ratio $\lambda=L(t)/L_0(t)$ (also shown in Table \ref{Twidth})
 is always greater than 1 and  {\em does not depend on time\/}:
 \begin{eqnarray}\label{LL0}
&& \lambda=\frat{L(t)}{L_0(t)}=\frat{2}{\alpha^2}
  \left[\exp(\alpha)-1-\alpha\right]= \nonumber \\
&&  =1+\frat{\alpha}{3}+
   \frat{\alpha^2}{12}+\ldots\; ,\quad \frat{R}{c} <t<\infty\, .
\end{eqnarray}
  Thus, the time-spread luminosity ``tail" turns out to be
  brighter than that of the intrinsic luminosity $L_0(t)$ by
  a factor of $\lambda(\alpha)$. For $\alpha\gsim 2$
  this effect increases the apparent brightness by about one stellar
  magnitude or more.

  In the case of a power-law light curve,
  $L_0(t)\sim\left(1 +\frat{t}{\tau}\right)^{-n}$,
  that also allows the integral to be taken in a closed form,
  such an effect is absent: $L(t)/L_0(t)\rightarrow 1$
  for $t\rightarrow\infty$. The exponential function decreases
  fast enough to make the luminosity tail remain ever overluminous.

\section{Time Delay Relativistic Effects}

There exist a number of constraints on the lower limit of integration
$\mu_{\mathrm{min}}$ in Eq. (\ref{far_transport}) and, accordingly,
there are several reasons why the light from the source does
not reach the observer.
\begin{enumerate}
\item  The first constraint is a geometric one determined
by the angular sizes of the radiation source
visible to the observer. For example, the back of the
surface of the spherical envelope at rest is disregarded
in the integration, which corresponds to a zero lower
limit of integration for any time (Fig.~\ref{geom3}a).
\item  The second constraint is a dynamical one related
to the finiteness of the speed of light.
If the radius $R(t)$, where radiation begins to freely propagate
outwards, does not depend on or shrinks with time then
$\mu_{\mathrm{min}}=0$. However when $R(t)$ increases with time
the lower limit of integration becomes
$\mu_{\mathrm{min}}\equiv\beta_0 =\frac{1}{c}\, dR(t)/dt$ (Fig.~\ref{geom3}b).
%For this reason, the cosine of the angle to the surface of the
%sphere $\mu$ can not be smaller than the velocity of
%the outer edge of the expanding envelope, i.e., $\mu > \beta_0$
%(Fig.~\ref{geom3}b), and the lower limit of integration becomes
%$\beta_0$.
\item  An even more severe constraint can be in the
case of accelerated envelope expansion. In this case,
it can reach some spatial point faster than the light
emitted from the edge (relative to the observer) regions
of the envelope (Fig.~\ref{geom3}).
%The condition on the
%cosine of the angle for the lower limit of integration
%then becomes
%\[
%        R^2(t)+(cdt)^2+2R(t)cdt\mu<R^2(t+dt),  \forall dt>0
%\]
%This condition can be easily represented geometrically:
%as if the radiation comes to the observer from
%an opaque body bounded by the equitemporal surface
%from which the photons reach the remote observer
%simultaneously (Fig.~\ref{geom3}d).
% ALTO NEW BEGIN
The light from the envelope point $(R(t),\mu)$ does not reaches the observer in case
 the following condition is performed:
\begin{eqnarray}
   &&  \exists dt: G(t,\mu,dt) = \nonumber \\
   &&  = R^2(t)+(cdt)^2+2R(t)\mu cdt - R^2(t+dt) < 0\, .
\end{eqnarray}
After all these points are determined the flux for the distant observer can be
found by integration over the equitemporal surface
from which the photons reach the distant observer simultaneously and
by excluding all found points from the integration (Fig.~\ref{geom3}d).
This constraint is not active if second derivative of the function $G(t,\mu,dt)$ is
always positive.
%
%This constraint is not active if second derivative of the function $G(t,\mu,dt)$ is
%always positive, i.e.:
%\begin{equation}
%    1-B(t)^2-\frac{R(t)}{c}\frac{dB(t)}{dt}>0\, ,
%\end{equation}
%where $B(t)=dR(t)/dt$. Thus the light curve should be affected significantly
%in case of large and highly accelerated envelopes.
% ALTO NEW END
\end{enumerate}

\begin{figure}
\centering
\includegraphics[width=84mm]{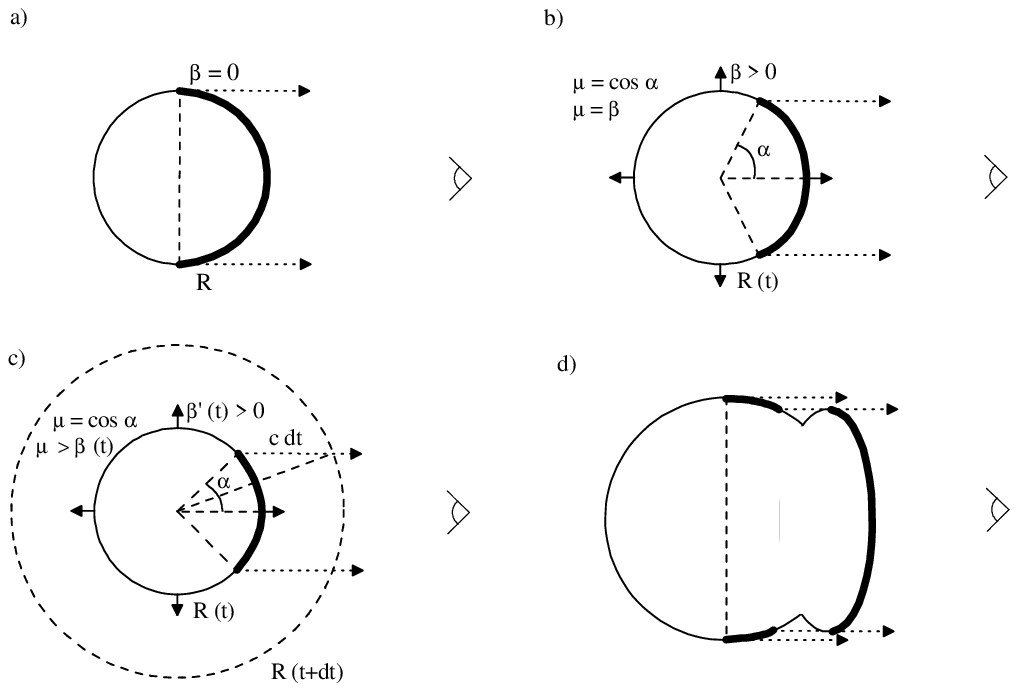}
\caption{Dependence of the radiation flux at a remote point of observation
        (on the right) on the dynamical characteristics of
        the observed envelope. The envelope region from which the radiation
        reaches the observer is highlighted by the heavy line for
        the following cases: the envelope is at rest (a),
        expands with a constant velocity $v=\beta c$ (b),
        and expands with acceleration (c). Panel (d) shows the equitemporal
        surface the radiation from some regions of which does not reach the observer due to the
        peculiar envelope dynamics.
}
\label{geom3}
\end{figure}

The last constraint is most stringent and includes
the previous constraints as special cases. Let us
demonstrate this using the following example. For
the dynamics of an expanding envelope with a linear
increase in velocity, the propagation law can be
written as $R/c = R_0+\beta_0 t + at^2$, where $R_0$, $\beta_0$, and
$a$ are constants. Generally, the analytical solution is
cumbersome and cuts out certain intervals of angles
from the flux integration. However, it is reduced to
$\mu_{\mathrm{min}} = \beta_0+2at/c$ for $R_0=0$ and we can see that, in
this case, the flux will be lower than that in the case of
envelope motion with a constant velocity.

To demonstrate the peculiarities of the flux behavior
at the point of observation, let us consider the
effect of a flux decrease in the problem of an outburst
of radiation on the envelope surface. This problem for
an instantaneous outburst and an envelope at rest
was solved by \citet*{Imshennik1981}. It is similar to
the problem of a light echo from a nova outburst considered
long ago by \citet{Couderc1939}, see also
\citet{BarkovBisnovatyiKogan2005} for an infrared outburst
generated by a gamma-ray burst in a dust cloud. In
contrast, we investigate the influence of an outburst
with duration $dt$ followed by envelope expansion on
the radiation flux at the point of observation; the
source cannot be considered stationary and instantaneous
and the envelope motion can be relativistic.

Consider a spherical envelope with radius $R_0=c$
on the surface of which an outburst with intensity
$I_{\nu}=100I_{\nu,0}$ and duration $dt=0.1$ s occurred at $t=0$
followed by envelope expansion according to the law
$R = R_0+0.5c(t-dt)$. The calculated radiation flux
from the envelope at a remote point of observation
is shown in Fig.$\,$\ref{geom5} (scenario $1$). Also shown here is
a comparison with the flux calculated for the case
where only the outburst without subsequent motion
occurred on the envelope surface (scenario $2$) and
the flux calculated when there was no outburst but
the expansion followed (scenario $3$). The gentler flux
decline in the first scenario than that in the second one
stems from the fact that the flux from the envelope
starting to move at $t=0.1$ s is higher that from the
envelope at rest. The steep decline near $t=0.6$ s is
related to the fact that the light from the moving parts
of the envelope arrives at the point of observation
earlier than that from the parts at rest and by $t=0.8$ s
the moving parts completely obscure the delayed flux
of the outburst from the envelope edges and the flux
behavior begins to follow scenario $3$.

\begin{figure}
\centering
\includegraphics[width=84mm]{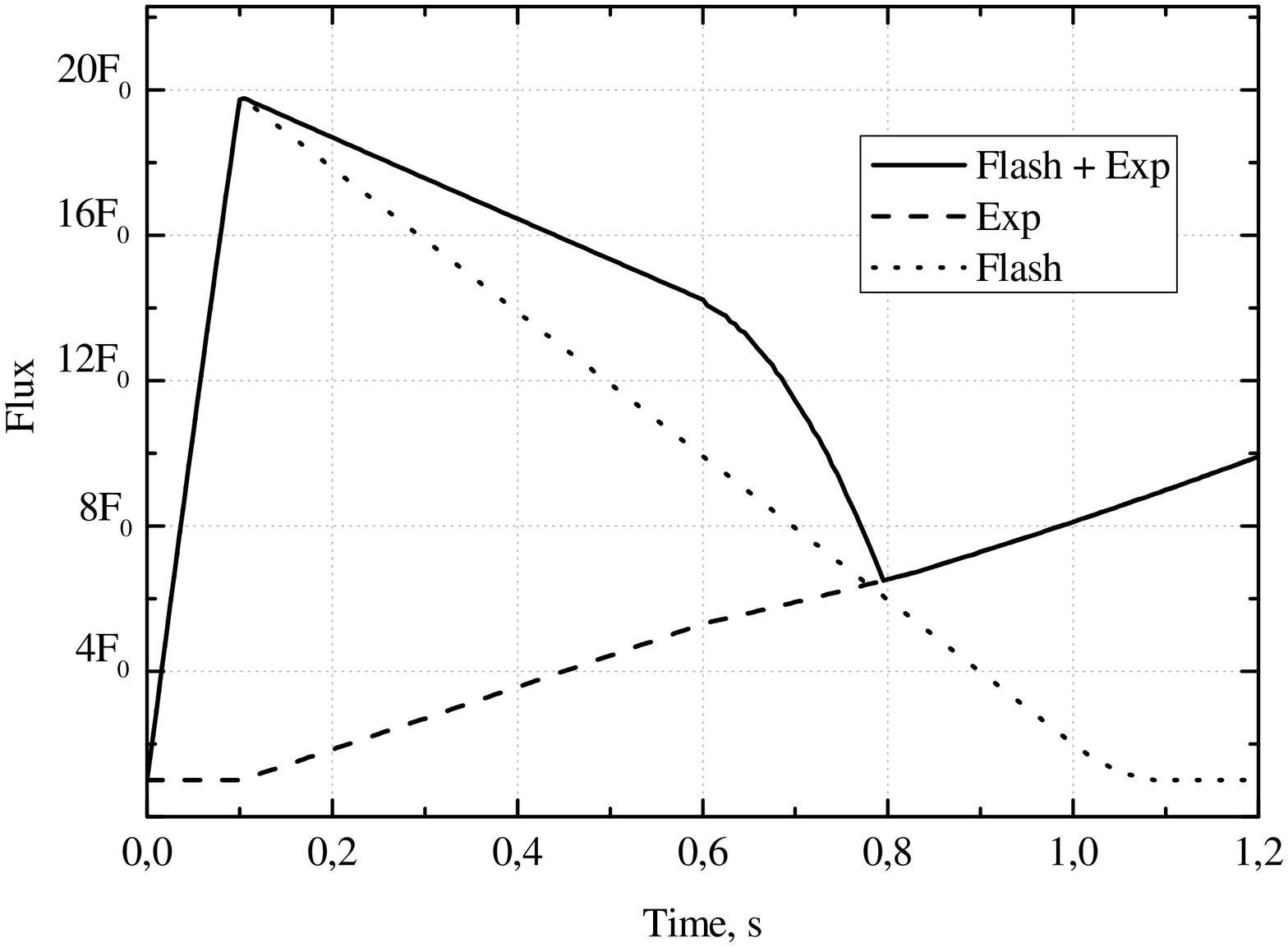}
\caption{Radiation flux at a remote point of observation from an envelope
         of radius $R_0=c$ on the surface of which an
         outburst with intensity $I_{\nu}=100I_{\nu,0}$ and duration $dt=0.1R_0 / c$ s
         occurred at $t=0$ followed by
         envelope expansion $R = R_0+0.5c(t-dt)$ (scenario 1). The dotted line denotes
         the flux as if only the outburst without any subsequent expansion
         occurred (scenario 2) and the dashed line represents only the motion (scenario 3).
         A zero time of observation corresponds to
         the outburst onset; $F_0$ is the radiation flux at a remote point of observation
         at the initial time $t=0$.
}
\label{geom5}
\end{figure}

%\end{verbatim}
\label{lastpage}
\end{document}